\documentclass[useAMS,usenatbib]{mn2e}

\usepackage{graphicx}

\title[Radio Observations of GRB Hosts]{Radio Observations of GRB Host Galaxies}
\author[E.~R.~Stanway, A.~J.~Levan \& L.~J.~M.~Davies]{Elizabeth R. Stanway$^{1}$\thanks{E-mail:
e.r.stanway@warwick.ac.uk}, Andrew J.~Levan$^{1}$ and Luke J.~M.~Davies$^{2,3}$\\
$^{1}$Department of Physics, University of Warwick, Gibbet Hill Road, Coventry, CV4 7AL, UK\\
$^{2}$H.~H.~Wills Physics Laboratory, University of Bristol, Tyndall Avenue, Bristol, BS8 1TL, UK\\
$^{3}$ICRAR, The University of Western Australia, 35 Stirling Highway, Crawley, WA 6009, Australia}

\begin{document}

\date{}

\pagerange{\pageref{firstpage}--\pageref{lastpage}} \pubyear{2014}

\maketitle

\label{firstpage}

\begin{abstract}
We present 5.5 and 9.0\,GHz observations of a sample of
seventeen GRB host galaxies at $0.5<z<1.4$, using the radio continuum to explore
their star formation properties in the context of the small but
growing sample of galaxies with similar observations. Four sources are
detected, one of those (GRB\,100418A) likely due to lingering
afterglow emission. We suggest that the previously-reported
radio afterglow of GRB\,100621A may instead be due to host galaxy
flux. We see no strong evidence for redshift evolution in the typical
star formation rate of GRB hosts, but note that the fraction of `dark' bursts with
detections is higher than would be expected given constraints on the
more typical long GRB population. We also determine the average
radio-derived star formation rates of core collapse supernovae at
comparable redshift, and show that these are still well below the
limits obtained for GRB hosts, and show evidence for a rise in
typical star formation rate with redshift in supernova hosts.

\end{abstract}


\section{Introduction}\label{sec:intro}

Long Gamma Ray Bursts (GRBs) are potentially a valuable probe of star
forming galaxies. Arising from the core collapse of a massive star \citep[e.g.][]{2006ApJ...637..914W},
they indicate the presence of recent star formation and are typically
associated with the most intensely star forming region of their host
galaxies \citep{2006Natur.441..463F,2010MNRAS.tmp..479S}. Optical and ultraviolet observations of those hosts galaxies
have suggested that these tend to be relatively low in metallicity
\citep[e.g.][]{2013ApJ...774..119G} and mass
\citep[e.g.][]{2010ApJ...721.1919C}, with modest star formation rates
\citep[e.g.][]{2004A&A...425..913C,2009ApJ...691..182S}. Thus they allow the selection and
study of a galaxy population that may represent the low mass tail of
the ultraviolet-selected, high redshift Lyman break galaxy luminosity function, and
could plausibly dominate the ionising photon density of the
Universe at early times \citep{2009ApJ...691..152C,2012ApJ...754...46T}.

If GRBs are an unbiased tracer of star formation, then they would be
expected to occur in the galaxies dominating the star forming
population at a given redshift. The phenomenon of `downsizing'
\citep{1996AJ....112..839C} describes the tendency of star forming
galaxies at low redshifts to be lower in mass, and also in
metallicity, than the bulk of the galaxy population. Local radio
surveys suggest that the majority of star formation at $z<0.1$ occurs
in a narrow range of radio luminosities, corresponding to star
formation rates of 1.5-15\,M$_\odot$\,yr$^{-1}$
\citep{2002AJ....124..675C}.  More massive galaxies formed their stars
at earlier times, typically at $z>1$, and often at a higher specific star
formation rate
than is seen in the local Universe \citep[see, for example,][and
  references therein]{2004ApJ...603L..69C,2010MNRAS.402.1693H}.  Thus
the peak in the volume-averaged cosmic star formation rate, in AGN
activity (indicative of co-evolution of spheroids) and in the number
density of obscured starbursts such as submillimeter galaxies \citep{2005ApJ...622..772C} all
occur at $z>1$, and the mass and luminosity of a typical starburst
decrease sharply towards $z=0$.

However, there are indications that the host galaxies of GRBs do not
necessarily trace this evolving, typical star forming population, but rather
that the occurence of a GRB may be likely in galaxies with certain
physical properties regardless of redshift \citep[see][for a recent
  review]{2014PASP..126....1L}. Extensive work in the optical and
near-infrared has now produced a reasonably clear picture of the
stellar populations in GRB hosts, in those examples sufficiently close
to carry out a detailed analysis
\citep[e.g.][]{2009ApJ...691..182S,2010MNRAS.tmp..479S} and in samples
designed to have well understood completeness and selection effects 
\citep[e.g. the `TOUGH' sample,][]{2012ApJ...756..187H,2012ApJ...755...85M}. 

By fitting the spectral energy distribution of a sample of host
galaxies, \citet{2009ApJ...691..182S} found no evidence for evolution
in host galaxy specific star formation rate or metallicity in the
redshift interval $0<z<6$, although available metallicity information
was limited to damped Lyman-$\alpha$ system measurements at
$z>2$. \citet{2012ApJ...749...68S} did find evidence for a strong
evolution in either the luminosity or density distribution of GRB
hosts and attributed this to a metallicity dependence in the GRB
progenitor population.

While \citet{2010MNRAS.tmp..479S} did not consider redshift evolution,
they compared GRB host locations to those of the core-collapse
supernova population (which should also trace star formation) and
found that the former occupy galaxies with higher optical surface
brightness but less mass and smaller radii, as also suggested by
\citet{2006Natur.441..463F} and others. Work on supernovae themselves
also support this. GRBs have been observationally associated with type
Ic supernovae \citep{1998Natur.395..670G} but not with other types, and may
well trace only a part of the Ic population, with additional constraints
(e.g. on the progenitor rotation) necessary to generate a GRB. In
local supernova samples, SNe Ic occur in more metal rich,
exceptionally star forming galaxies, compared to the hosts of other
core collapse supernovae \citep{2012ApJ...759..107K}, strengthening
the case for GRBs picking out some subset of the full star forming
galaxy population.

However this conclusion is by no means undisputed. As
\citet{2013MNRAS.432.2141C} point out, a large number of observational
selection effects as well as astrophysical biases must be taken into
account.  Initial suggestions that the GRB host population may show a
sharp metallicity cut-off
\citep[e.g.][]{2012ApJ...749...68S,2009ApJ...702..377K} have been
moderated by the detection of higher metallicity systems
\citep[e.g.][]{2010AJ....140.1557L,2012arXiv1205.4036K} with a more
gradual metallicity-dependent fall off in GRB rate now suggested
\citep[e.g.][and references
  therein]{2011ApJ...735L...8K,2012ApJ...744...95R}. While some
recent studies conclude that GRBs show a pronounced aversion to 
high metallicity \citep[e.g.][]{2013ApJ...774..119G}, there are also
those who believe no such dependence exists. Studies of the GRB host
galaxy mass function and long GRB burst rate, based on a large sample observed by
the GROND team, have suggested that no obvious biases in the host galaxy
properties with metallicity, mass or redshift
\citep{2012A&A...539A.113E} can be inferred, leaving this still very much an open
question.

So, if the degree to which the properties of GRB hosts track those
of the typical star forming galaxy at a given redshift is uncertain,
converting a gamma ray burst rate to a volume averaged star
formation density (and hence to the ionizing photon density vital for, amongst
other things, hydrogen and helium reionisation) is fraught with difficulties
and assumptions. Not only is the detection of bursts, their follow-up
and redshift confirmation, rather an erratic process, with a difficult
to establish completion function
\citep[see][]{2009ApJS..185..526F,2012ApJ...756..187H,2012arXiv1205.4036K},
but the derived star formation rates rely on prescriptions relating
the observed death of a single massive star in each host galaxy to the
stellar birth rate \citep[e.g.][]{2012ApJ...744...95R}.

Even studies to characterise GRB host samples in the optical are
subject to their own selection biases, such as the possibility that
some GRB host galaxies may be characterised by a significant quantity
of dust and obscured star formation and hence omitted. Early
sub-millimeter and radio observations suggested that as many as a
third of GRBs could arise in these dusty and relatively massive
systems
\citep{2003ApJ...588...99B,2006MNRAS.369.1189P,2004MNRAS.352.1073T}.
The definition of a population of `Dark GRBs' with low optical
emission relative to their X-ray flux \citep{2004ApJ...617L..21J}
again strengthened earlier suggestions that dust extinction may play a role that must be
understood in the interpretation of these sources
\citep[e.g. see discussion in][and references therein]{1998ApJ...493L..27G,2001ApJ...562..654D,2005ApJ...624..868R,2009ApJ...693.1484C,2011A&A...526A..30G}.
Including this `dark' population leads to determining a much higher
average dust content in the GRB host population than previously
believed, but is still insufficient to explain the early submillimeter
results \citep{2013ApJ...778..128P}.  Studies of the lowest redshift
bursts \citep{2010MNRAS.409L..74S} and other more recent work
\citep[e.g.][]{2014arXiv1402.4006H,2013ApJ...778..172P,2012ApJ...755...85M,2012ApJ...761L..32W}
has suggested that the typical GRB host galaxy does not in fact host
large quantities of extinguished star formation.

As a result, samples of GRBs with measured radio fluxes remain
small, with early work focused on detailed studies of individual targets
\citep[e.g.][]{2005PASJ...57..147K,2009ApJ...693..347M,2011MNRAS.410.1496S}.
Recent work by \citet{2012ApJ...755...85M} represented a substantial
improvement in the sample of GRB hosts observed in the radio, presenting 
observations for 22 $z<1$ sources, including those derived from the
`TOUGH' sample \citep{2012ApJ...756..187H} and sources compiled from the
literature. Of these, three were detected. While the constraints
placed on individual sources varied significantly from object to
object, a stacking analysis based on the median redshift of the sample
suggested that the typical undetected
source may be relatively low in star formation rate
(SFR$<$15\,M$_\odot$ yr$^{-1}$) and have dust
extinction ($A_{2800\mu m}<6.7$\,mag).

Recent improvements in correlator bandwidth, and hence sensitivity, at
the major radio observatories has opened up the potential to change
this, with observations reaching star formation rates within a factor of
a few of the optical and ultraviolet derived values for GRB hosts
 now achievable with a reasonable investment of time.  In this
paper we extend the radio analysis of long GRB hosts to higher
redshifts, and increase the number of observed sources, aiming to
characterise their star formation without assumptions regarding their
redshift evolution. We use the radio continuum to explore the star
formation rates of the GRB host galaxies in our sample, in the context
of the small but growing number of galaxies with similar
observations. We present 5.5 and 9.0\,GHz observations of seventeen
GRB host galaxies at $z=0.5-1.4$. Of these, radio observations of
twelve sources are reported here for the first time. Of the
remaining sources, four have previously been observed at
$\sim$1.4\,GHz \citep{2012ApJ...755...85M}. We report 5.5\,GHz flux
constraints and significantly improve the observed star formation rate
limits in two cases, and present a tentative (2.3\,$\sigma$) detection
at 5.5\,GHz of a third source. For the final source we report a flux
limit consistent with the detection reported by
\citet{2013ApJ...778..172P}.

In section \ref{sec:obs} we introduce observations, taken at both the
Australia Telescope Compact Array (ATCA) and the Jansky Very Large
Array (VLA). In section \ref{sec:afterglow} we examine the two most
recent bursts in detail and establish interpretation as afterglow or
star formation. In section \ref{sec:SFRs} we discuss the conversion
from radio flux to star formation rate. In section
\ref{sec:discussion} we discuss our results in the context of previous
work in this field, before presenting our conclusions in section
\ref{sec:conc}.

Throughout, optical magnitudes are presented in the AB system. Host
galaxies are typically referred to by the name of the associated GRB,
but measurements are those of the host galaxy unless otherwise
specified. Where necessary, we use a standard $\Lambda$CDM cosmology
with $H_0=$70\,km\,s$^{-1}$\,Mpc$^{-1}$, $\Omega_M=0.3$ and
$\Omega_\Lambda=0.7$.

\section{Observations}\label{sec:obs}

 \subsection{Sample Selection}\label{sec:sample}

 Gamma ray bursts were selected from the data compilation
 table\footnote{http://swift.gsfc.nasa.gov/archive/grb\_table/}
 recording burst triggers from the {\em Swift} telescope
 \citep{2004ApJ...611.1005G}.  The primary criterion for sample
 selection was the existence of a known redshift for the GRB or its
 host galaxy, with sufficient precision to accurately
 derive rest-frame properties, and sufficiently low to allow useful
 radio constraints.

 As mentioned below and discussed in section \ref{sec:dark}, one initial goal of our
 ATCA programme (see
 section \ref{sec:obs_atca}) was to evaluate whether `dark bursts' differed in their
 radio properties from the wider population. As a result four of
 the sources in this study were identified as dark bursts (lying at $z=0.5-0.8$), and other long GRBs at similar
 redshifts were preferentially selected for our ATCA observations  so as to allow more direct comparison of
 their properties. The targets for our VLA observations
 (section~\ref{sec:obs_vla}) were chosen to lie at slightly higher redshifts, ideally at $z=1.0-1.4$,
 to extend this work to earlier times and take best advantage of the higher sensitivity of the VLA.

 Southern sources were prioritized so as to maximize coverage of the
 $uv$-plane in interferometric radio telescopes. If multiple targets
 satisfied these selection criteria and were accessible to the
 telescope during the observing programme, then the sources observed
 were selected at random.  As a result, and given the patchy
 spectroscopic follow-up of bursts (particularly in the early years of
 the {\em Swift} mission), the sample is unlikely to be complete, but
 nonetheless samples the known population. All sources all satisfy the
 T$_{90}>2.0$\,s criterion for a `long' duration gamma-ray burst
 \citep[where T$_{90}$ is the interval within which 90\% of the
   integrated counts from the burst are
   detected][]{1993ApJ...413L.101K}.

Our primary source for redshift information was the compilation
provided in the {\em Swift} burst lookup
table\footnote{http://swift.gsfc.nasa.gov/archive/grb\_table/}. This
table is hosted by Swift and seeks to compile data on these triggers
and their follow up from GCN announcement and other sources, although
its data is occassionally superceded by more detailed analyses. For
two sources in our sample, GRB\,060814 and GRB\,071003, two highly
discrepant redshifts appear in the literature.

 GRB\,071003 was initially reported as lying at $z=1.10$ due to the
 presence of an absorption line system in the afterglow at this
 redshift \citep{2007GCN..6850....1P}. However reexamination of the
 afterglow spectrum revealed the presence of a weaker but convincingly
 detected line system at $z=1.604$ \citep{2008ApJ...688..470P}.  This
 latter redshift is now the accepted value for this burst. While we initially
selected this source for observation based on the lower redshift, we adopt the
 better supported value of $z=1.60$ for the analysis in this paper.

 GRB\,060814 was selected for follow-up based on the $z=0.84$
 interpretation given in the {\em Swift} burst table, derived from
 Keck spectroscopy \citep{2007GCN..6663....1T} at early times and
 appearing in several catalog papers
 \citep[e.g.][]{2010ApJ...720.1513K}.  However later observations have
 suggested that the faint host galaxy was misidentified and a second
 redshift $z=1.92$ was reported, from line detections in a star
 forming galaxy determined to be closer to the burst location
 \citep{2012ApJ...749...68S,2012arXiv1205.4036K}. Throughout, we give
 results for the $z=1.92$ hypothesis based on the higher probability
 that this is the host galaxy redshift. We note that the measured flux
 limit would correspond to a substantially lower star formation rate
 if $z=0.84$ is used instead (as figure \ref{fig:radio_results}
 illustrates). If this target were detected, it would also be subject
 to some uncertainty as to whether the flux arose from the GRB host,
 or the intervening $z=0.84$ galaxy. Due to the likely high redshift
 (which renders comparison with the rest of our sample problematic)
 and degree of ambiguity, this source is omited from statistical
 analyses in section \ref{sec:discussion}.

 \subsection{ATCA Observations}\label{sec:obs_atca}

  Radio continuum observations of 13 GRB host galaxies were undertaken
  in 2011 April 15-19 using the Australia Telescope Compact Array
  (ATCA)\footnote{Observations associated with programme C2544}. The targets for this programme were selected as
  {\em Swift}-detected long gamma-ray bursts lying at $z\sim0.5-0.8$, based
  on published observations. Four targets were included which satisfy
  the conditions for being defined as `Dark GRBs' (i.e. optically
  faint, see section \ref{sec:dark}).

  Data were taken simultaneously at 5.5 and 9.0\,GHz, with
  a correlator bandwidth of 2\,GHz centered on each frequency. The
  telescope was in its most elongated 6A configuration, with maximum
  and minimum baselines of 5.938 and 0.337\,km respectively, aligned
  East-West, and earth rotation synthesis was used to improve coverage of the
  $uv$-plane. Data were taken across the full range of possible hour
  angles, with a total on-source integration time per target of
  $\sim$140 minutes. A bright, compact source close on the sky to each
  target was used for phase calibration, and absolute flux and
  bandpass calibration were determined through observations of PKS
  1934-638 (the standard calibrator for ATCA).

  Data were reduced using the standard software package {\sc miriad}
  \citep{1995ASPC...77..433S}, applying appropriate bandpass, phase
  and flux calibrations, and after flagging the dataset for radio
  frequency interference. Each band comprised 2048 channels, each of
  1\,MHz bandwidth. Multi-frequency synthesis images were constructed
  using natural weighting and the full bandwidth between the flagged edges of each band. The
  resulting synthesised beam depends on
  $uv$-plane coverage. This tends to be relatively poor for northern
  sources, so the beam is elongated in declination as given in
  table \ref{tab:atca_obs}. The targets were placed close to the
  centre of the 8.5\,arcmin primary beam. Given their redshifts and
  the typical identification of GRB hosts as relatively low mass
  galaxies \citep[e.g.][]{2010MNRAS.tmp..479S,2009ApJ...691..182S} our GRB host targets
  were expected to be compact or point sources in these
  observations. In each image, multiple faint sources were detected, and
  the images were `cleaned' using standard prescriptions.

  The resulting 5.5\,GHz synthesis images were inspected for any flux excess in
  close proximity to the burst or its host galaxy (where known). The
  typical uncertainty in the enhanced {\em Swift} XRT position for the burst \citep{2009MNRAS.397.1177E} is
  1.4 arcseconds, with a few cases where the constraints are poorer
  ($\sim$3 arcsec). In four fields, a flux excess exceeding the
  typical image noise by a factor of two were observed within the XRT
  error circle, or coincident with a known host galaxy. Either
  incomplete sampling of the $uv$-plane or the presence of a bright
  object close to the telescope beam can leads to flux 
  being scattered across a synthesised image in a
  correlated noise pattern. We inspected each target to ensure that
  there was no sign of large scale pattern noise that might explain 
  a flux excess at the target location.

  In order to rigourously determine the 5.5\,GHz flux (or limit) for each
  target and its associated uncertainty, we made use of the {\sc
    miriad} task `imfit', specifying that the software attempt to fit
  a point source, matching the dimensions of the synthesised beam.  In
  order to allow the most conservative limit on the host galaxy flux
  (i.e. the maximum value permitted by the data), we permitted the
  algorithm to search around the XRT position by up to twice the width
  of the elongated synthesised beam in Right Ascension and a beam
  width in declination, sufficient to encompass both the XRT error
  circle and any plausible uncertainty associated with bandwidth
  smearing or the telescope pointing algorithm. This typically
  resulted in a search box of $\sim10\times3''$, and a measured offset
  of peak flux from the provided coordinates of $<2$\,arcseconds. 

  For non-detections, we report the maximum permitted point source flux
  and its associated error (scaled by the software to account for correlated
  noise), based on this procedure in table \ref{tab:atca_obs}. These represent
  conservative limits. The scaled errors, which account for the difficulty in
  determining a flux excess above the structured background noise exceed the 
  image root-mean-square noise (often quoted as a flux uncertainty) by 
  a factor $>5$. 
  
\begin{figure*}
\includegraphics[width=0.75\columnwidth]{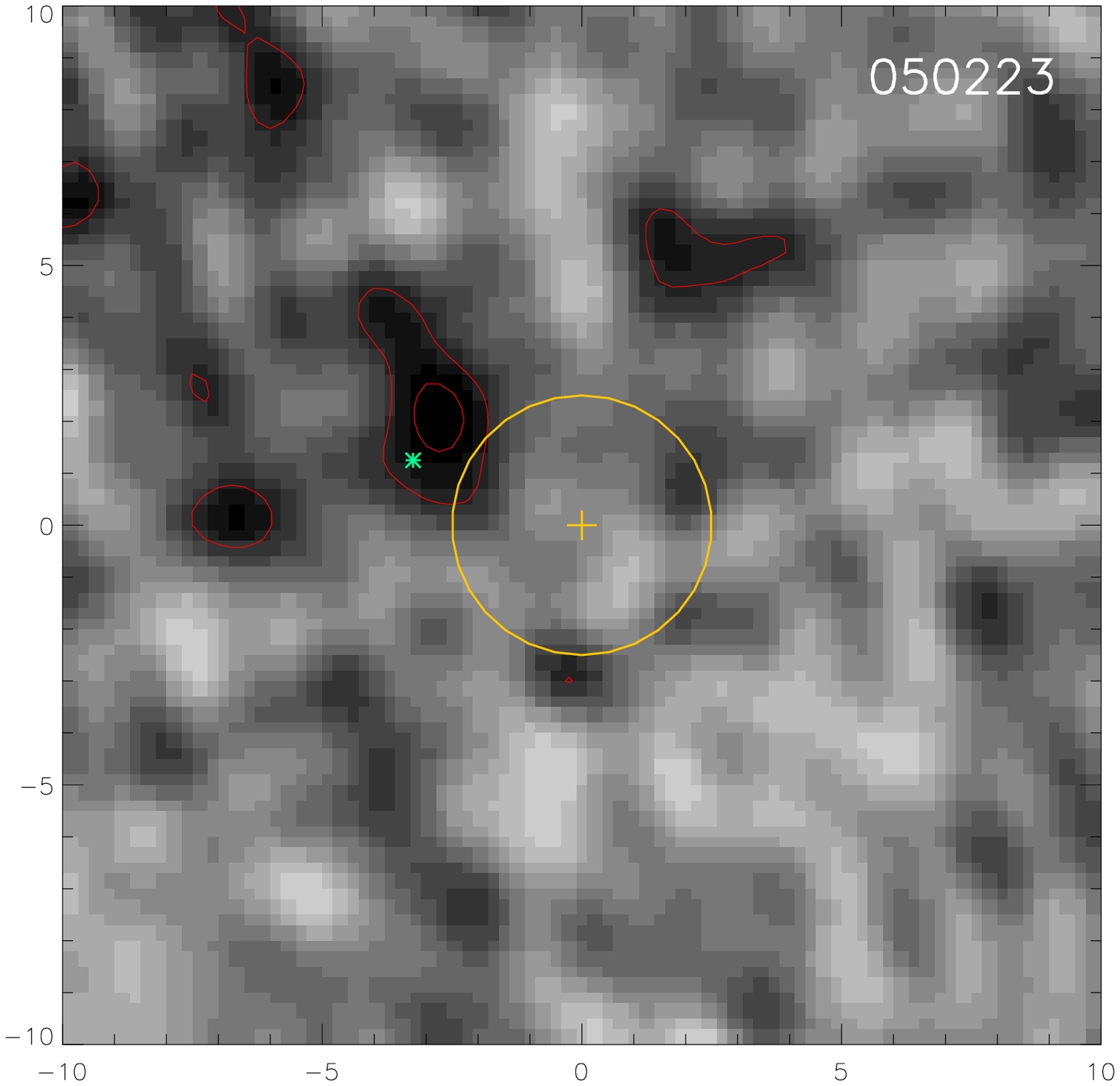}
\includegraphics[width=0.75\columnwidth]{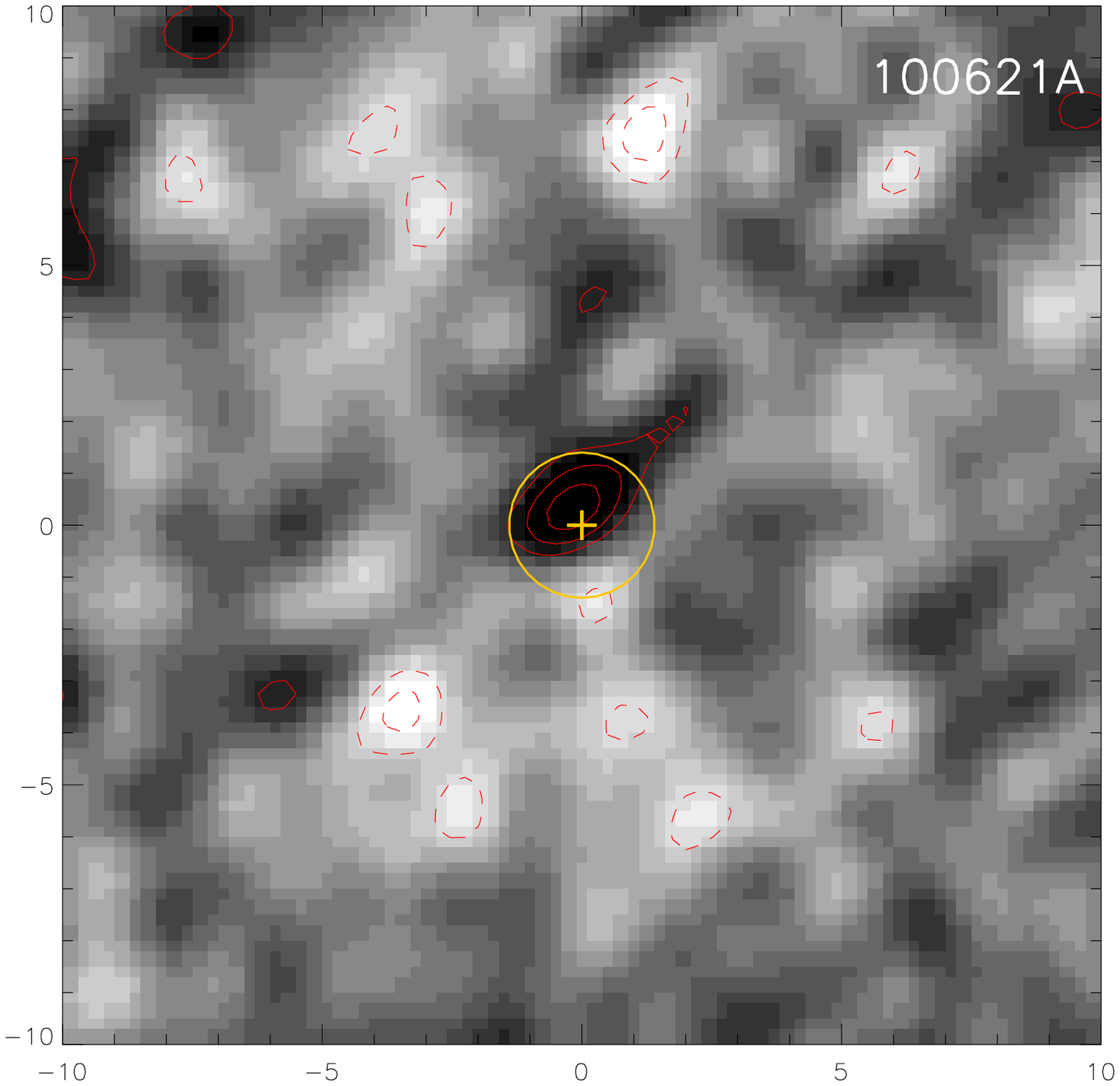}
\includegraphics[width=0.75\columnwidth]{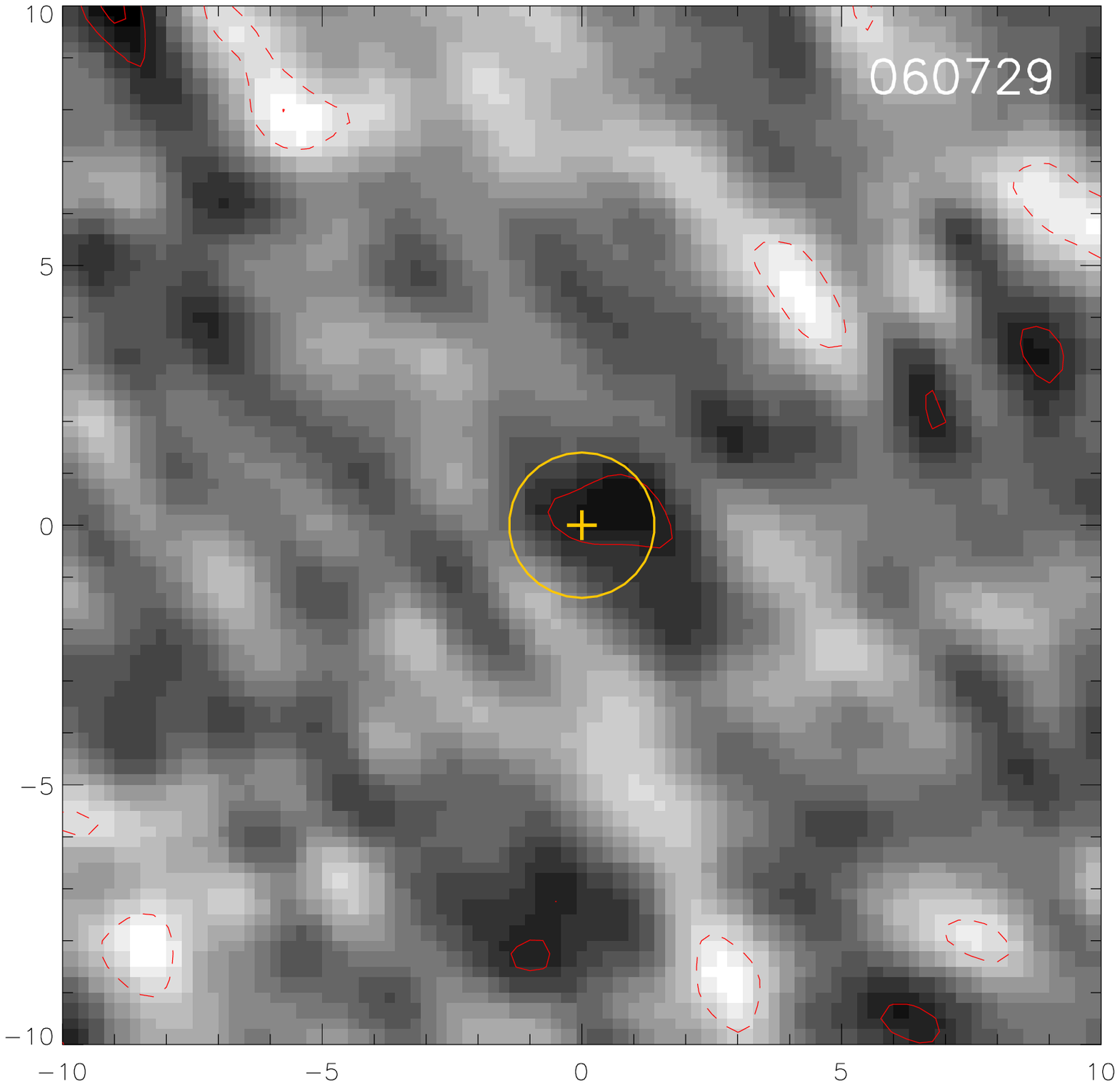}
\includegraphics[width=0.75\columnwidth]{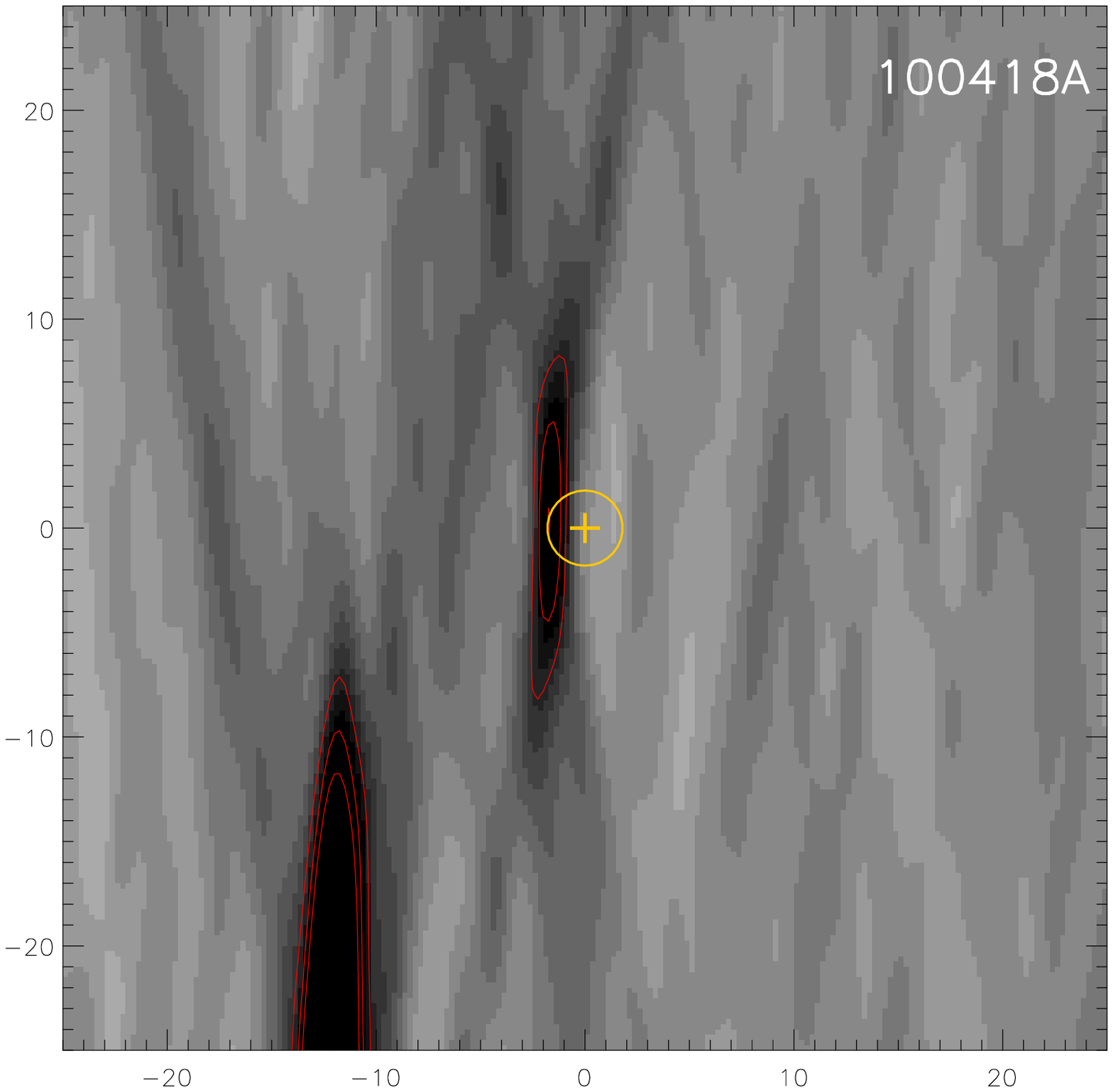}
\caption{5.5\,GHz radio maps of the four detections identified in table \ref{tab:atca_obs}. We note again that the detection of the host of GRB\,060729 is very tentative, at 2.3$\sigma$. Dark colours indicate higher fluxes, and contours are indicated at $\pm$2, 3 and 4 times the image noise level. The yellow cross and circle mark the {\em Swift} XRT location identified for the afterglow and its associated error circle. In the case of GRB\,050223, the second cross indicates the position of the host galaxy identified by \citet{2006A&A...459L...5P}, with which the radio detection appears to be associated. Axes are labelled in arcsecond offsets from the XRT location, with North up. Note that the synthesised beam for GRB\,100418A is very extended in declination, as shown in table  \ref{tab:atca_obs}.\label{fig:radio_maps}}
\end{figure*}

  For the four sources with an identified flux excess within the
  search region, we explored the possibility of fitting an extended
  source, resulting in a poorer fit to the data in all four cases. In
  each case, `imfit' returned a position consistent with the peak flux
  identified by visual inspection of the images. For consistency, and based
  on the large synthesised beam of the telescope at these frequencies, we
  report the point source flux for these objects together with the reported
  estimate of uncertainty in table \ref{tab:atca_obs}, and present 5.5\,GHz
  radio maps of these objects in figure \ref{fig:radio_maps}.

  Of these objects, two (GRBs\,100621A and 100418A) are well defined,
  clear radio detections coincident with the GRB X-ray location. A
  third (GRB\,050223) shows a significant (3\,$\sigma$) point source
  coincident with the host galaxy identified by
  \citep{2006A&A...459L...5P}. The fourth source (GRB\,060729)
  presents an excess of 2.3\,$\sigma$ over the background noise,
  coincident with the burst location.  

  While marginal as a claimed detection, this flux excess remains when
  the data is sub-divided, or a different weighting is used for image
  reconstruction. In addition to being coincident with the X-ray
  location, the flux excess is stronger than any of the likely noise
  features in its vicinity on the image, and is not straightforwardly
  attributable to the beam pattern extending from any other object.
  The probability of a noise fluctuation of this strength occuring in our
  search region by chance is 3\%. In our sample of 17 sources
  (combining ATCA targets with the VLA observations discussed in
  section \ref{sec:obs_vla}) we might expect 0.5 such fluctuations
  (assuming gaussian statistics). We note that no targets in our
  sample show a negative flux of comparable significance. Thus, while
  it remains possible that this radio flux is not attributable to the
  target, in the analysis that follows, we treat the measured 5.5\,GHz
  flux as a radio detection, with the caveat that it might also be
  treated as a robust upper limit.

\begin{table*}
\begin{tabular}{lccrcccccc}
Object ID & location  & z & 5.5\,GHz Flux & S/N & Beam size & SFR                 &  9.0\,GHz Flux \\ 
          &   (J2000) &   &  $\mu$Jy/beam &     &           & M$_\odot$\,yr$^{-1}$ & $\mu$Jy/beam \\ 
\hline\hline 
GRB\,081007   & 22:39:50.49 	-40:08:49.1  & 0.529 &   38.1 $\pm$ 26.7   &         &   3.9$\times$2.2$''$   &       $<$\,35   &  \\       
GRB\,060729   & 06:21:31.79 	-62:22:12.4  & 0.54  &   65.4 $\pm$ 27.8   &  2.3    &   4.1$\times$1.8$''$   &   55 $\pm$ 24   &  60 $\pm$ 41  \\      
GRB\,100621A  & 21:01:13.10 	-51:06:22.8  & 0.542 &  120.1 $\pm$ 31.9   &  3.8    &   3.5$\times$2.2$''$   &  101 $\pm$ 27   & 106 $\pm$ 42  \\      
GRB\,090424   & 12:38:05.12 	+16:50:15.4  & 0.544 &   36.6 $\pm$ 28.0   &         &    10$\times$1.7$''$   &       $<$\,38   &  \\       
GRB\,050223   & 18:05:32.35 	-62:28:19.7  & 0.591 &   90.5 $\pm$ 30.1   &  3.0    &   3.3$\times$2.0$''$   &   93 $\pm$ 31   &  93 $\pm$ 48  \\      
GRB\,050525A  & 18:32:32.64 	+26:20:21.6  & 0.606 &   15.6 $\pm$ 33.8   &         &   8.4$\times$1.6$''$   &       $<$\,53   &  \\       
GRB\,100418A  & 17:05:27.19 	+11:27:39.8  & 0.623 &  363.0 $\pm$ 48.0   &  7.6    &    18$\times$1.6$''$   &        ---      &  199 $\pm$ 57 \\      
GRB\,051022   & 23:56:04.11 	+19:36:23.7  & 0.809 &   23.0 $\pm$ 35.5   &         &    14$\times$1.5$''$   &       $<$\,98   &  \\       
GRB\,070508   & 20:51:11.80 	-78:23:05.1  & 0.82  &   35.0 $\pm$ 28.2   &         &   4.3$\times$1.6$''$   &      $<$\,101   &  \\       
GRB\,071112C  & 02:36:50.95 	+28:22:16.7  & 0.823 &   50.1 $\pm$ 25.2   &         &   7.3$\times$1.6$''$   &      $<$\,126   &  \\       
GRB\,050824   & 00:48:56.20 	+22:36:32.9  & 0.83  &   42.3 $\pm$ 33.2   &         &    10$\times$1.5$''$   &      $<$\,111   &  \\       
GRB\,080710   & 00:33:05.63 	+19:30:05.5  & 0.845 &   42.6 $\pm$ 28.8   &         &    11$\times$1.6$''$   &      $<$\,112   &  \\       
GRB\,060814   & 14:45:21.32 	+20:35:09.2  & 1.92  &   43.6 $\pm$ 23.5   &         &   9.1$\times$1.6$''$   &      $<$\,670   &  \\       
GRB\,091208B  & 01:57:34.10 	+16:53:22.9  & 1.063 &    0.0 $\pm$ 4.2    &         &   1.3$\times$1.1$''$   & $<$\,33   &  \\     
GRB\,080413B  & 21:44:34.60 	-19:58:51.8  & 1.101 &    7.6 $\pm$ 4.7    &         &   1.9$\times$1.0$''$   & $<$\,39   &  \\      
GRB\,100901A  & 01:49:03.41     +22:45:30.1  & 1.408 &    0.2 $\pm$ 2.9    &         &   1.2$\times$1.1$''$   & $<$\,42   &  \\     
GRB\,071003   & 20:07:24.11 	+10:56:51.1  & 1.604 &    2.1 $\pm$ 4.3    &         &   1.4$\times$0.9$''$   & $<$\,83   &   \\

\end{tabular}
\caption{Results from the radio observations at 5.5 and 9.0\,GHz,
  taken at the ATCA and VLA. Objects are ordered by redshift, with the
  exception of GRB\,060814 (see section \ref{sec:afterglow}). Fluxes
  and 1\,$\sigma$ uncertainties are given in $\mu$Jy/beam. The beam
  size is given at 5.5\,GHz and is half this size at 9.0\,GHz. For
  measurements with an associated S/N$>2$, the signal to noise is
  shown in the fourth column. The penultimate column gives the
  inferred star formation rate in solar masses per year, assuming the
  radio flux is driven by star formation, as described in section
  \ref{sec:SFRs}, with 2\,$\sigma$ limits where appropriate. The
  detection of GRB\,100418A is not interpreted as star formation (see
  section \ref{sec:afterglow}). If GRB\,060814 is instead placed at
  $z=0.84$ (see section \ref{sec:sample}), the limit instead
  corresponds to $\sim107$\,M$_\odot$ yr$^{-1}$. For sources detected
  at 5.5\,GHz, the final column gives the observed-frame 9.0\,GHz flux
  in $\mu$Jy. No other sources are detected at 9.0\,GHz. The final
  four sources were observed at the VLA, and 9.0\,GHz data was not
  taken. Locations are given for the GRB afterglow and taken from the
  {\em Swift} XRT enhanced positions
  catalogue.\label{tab:atca_obs}\label{tab:vla_obs} }
\end{table*}

  Observations at 9.0\,GHz are more sensitive to atmospheric
  conditions and yield less sensitive constraints on star formation
  rate for all but the most peculiar of radio spectral slopes. We
  nonetheless inspect our 9.0\,GHz for targets coincident with the GRB
  locations, using the same procedure described above.  Two of the
  5.5\,GHz-detected sources were also detected at 9.0\,GHz.  The
  remaining sources showed no evidence for a 9.0\,GHz detection.
  
  We note that the constraints obtained above may be weakened if the
  sources are extended relative to the synthesised beam. The median
  effective radius of GRB host galaxies is $\sim1.7$\,kpc
  \citep{2007ApJ...657..367W}. This corresponds to a projected size of
  just 0.3$''$ at $z=0.5$ and 0.2$''$ at $z=1.0$. Of 47 GRB hosts in
  the \citet{2007ApJ...657..367W} sample, only one (GRB\,011121 at
  $z=0.34$) would be comparable to a 1.5$''$ beam size in our
  imaging. While it is possible that a more extended host may exist
  within our relatively small sample, such studies of archival
  space-based imaging suggest that this is unlikely.

 \subsection{VLA Observations}\label{sec:obs_vla}

  The survey was extended to higher redshifts ($z=1.0-1.6$) through the
  observation of four additional sources at the Karl G. Jansky Very
  Large Array (VLA). Observations were obtained in service mode in
  June-July 2012 and were associated with programme 12A-279 (PI:
  Stanway). A total bandwidth of 2\,GHz was centred at 5.5\,GHz, with
  data collected in 1024 channels, each 2\,MHz in width. Observations
  were performed in B configuration.

  Three targets - GRBs 071003, 080413B and 091208B - were each
  observed for $\sim$80 minutes on source, and a fourth target - GRB
  100901A - for $\sim$160 minutes.  As in the case of the ATCA
  observations above, a nearby point source was used for phase
  calibration on each target and absolute flux calibration was
  provided by the standard calibrator 3C48. Data were flagged and
  reduced using standard {\sc CASA} tasks, producing a multi-frequency
  synthesis image for each target, using natural weighting. Both the
  improved instantaneous $uv$-plane coverage and the northern location
  of the VLA relative to the ATCA led to a more compact synthesised
  beam in these observations, as shown in table \ref{tab:vla_obs}.

\begin{figure*}
  \includegraphics[width=1.8\columnwidth]{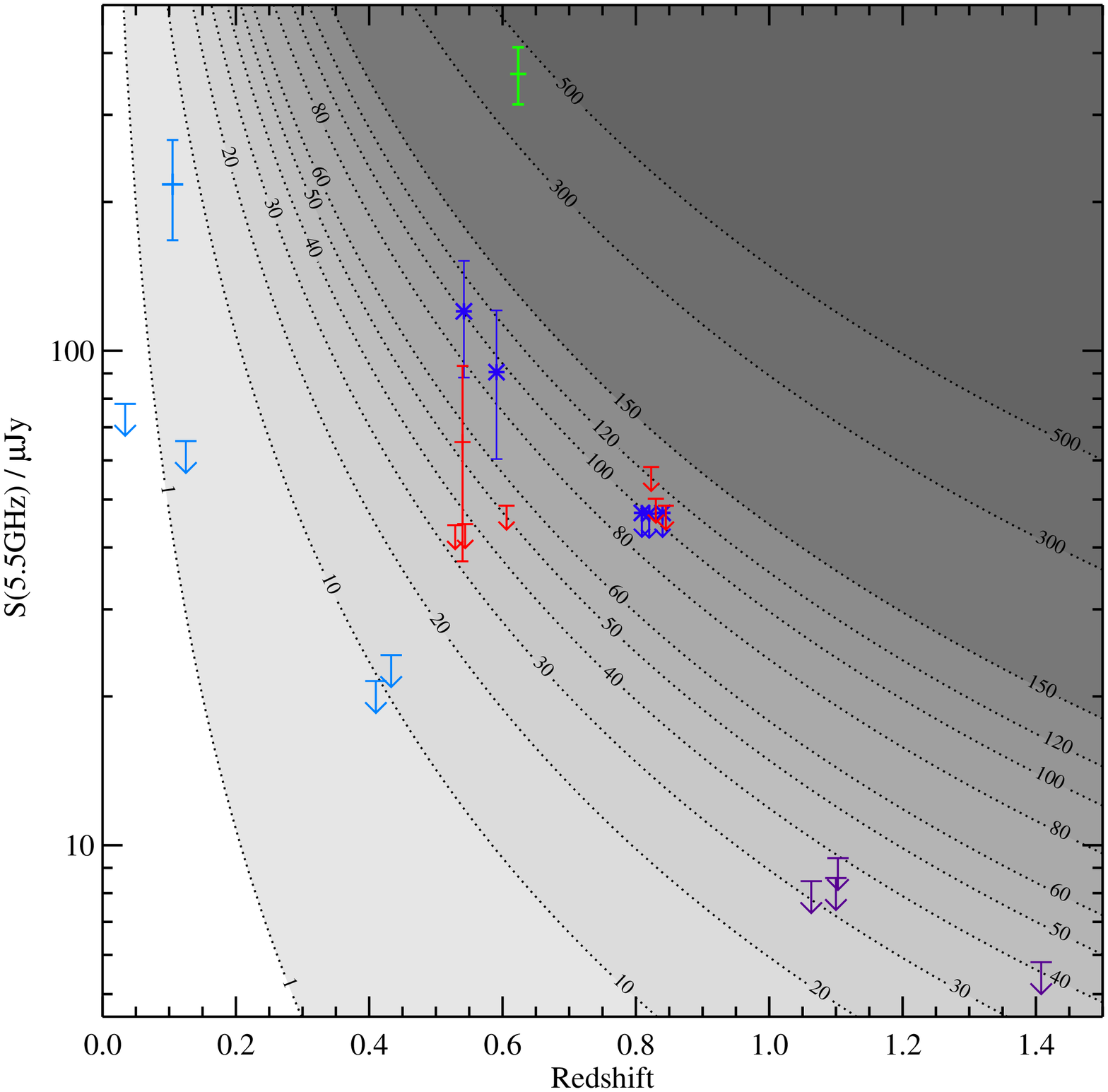}
\caption{The measured fluxes (or 2\,$\sigma$ limits thereon) of the
  samples discussed in this paper. The four points at $z>1$ are
  derived from VLA data, the remaining points from ATCA data. The five
  points at $z<0.5$ are those presented in Stanway et al (2010) and
  are shown to provide a point of comparison. The five observations
  targeting the location of `dark' bursts are indicated by asterisks
  (two detections, three limits). The contours indicate inferred
  radio-continuum-derived star formation rates in solar masses per
  year, as discussed in section \ref{sec:SFRs}. We note that the
  highest flux detection, that of GRB\,100418A, should not be
  interpreted in terms of star formation as discussed in section
  \ref{sec:afterglow}. GRB\,060814, at $z=1.92$, is omitted from this
  figure for clarity.\label{fig:radio_results}}
\end{figure*}

  Each image revealed a number of faint radio sources, as well as
  known catalogue sources.  In the field of GRB 071003, we recover the
  known NVSS galaxy J200658+110024 at a separation of 7.3\,arcmin from our
  target. As before, we permit a small search region, allowing for the 
  X-ray position uncertainty, any pointing uncertainty or offsets between
  the GRB or its host. None of the four targeted sources were individually
  detected. Given the non-detection, we measure the flux within the GRB error 
  circle, allowing for small ($<2''$) shifts in the centroid if these
  maximise the flux, thus achieving the most conservative constraints.  Our
  resulting  sensitive flux limits as shown in table \ref{tab:vla_obs}
  and figure \ref{fig:radio_results}.

\section{Afterglow or Host Galaxy?}\label{sec:afterglow}

The majority of the targets in this survey were observed a substantial
time ($>$2\,years) after the initial burst. However, in two cases -
GRBs 100418A and 100621A - radio fluxes were obtained less than a year
post-burst in the galaxy rest-frame. Given that these sources are two
of only four detections in our sample, the question arises: are we
observing host galaxy flux, or are our observations contaminated by
the late time radio afterglow of the burst itself?

We explore this possibility through archival imaging and the
literature. It has been established that GRB 100621A was unusually
bright. \citet{2013A&A...560A..70G} analysed the multi-wavelength
afterglow properties, including observations taken with ATCA in
Jun-Jul 2010 (4-27 days after burst) at the same frequencies as those
observed here.
In table \ref{tab:100621a} we report the fluxes measured by
\citet{2013A&A...560A..70G} and compare with those measured in our
observations. \citeauthor{2013A&A...560A..70G} suggested that
their 2010 Jul 17 observations (measured at 17 days post-burst
in the rest frame) might suggest a rapid fall off of the
afterglow, or that the early time data might be explained by
scintillation.

The radio afterglow of long GRBs can be observed across an extended period 
of time. Unlike the optical afterglow, which typically fades on a time scale
of hours, radio afterglows extend across days or months. The timescales
for this afterglow depend on the observed frequency, since high frequency flux
peaks earlier than low frequency flux.
 \citet{2012ApJ...746..156C}
compiled data on 95 GRBs for which there were radio detections spanning
from minutes to $\sim$500 days following the {\em Swift} gamma-ray trigger.
In 63 cases there was sufficient data to determine a peak flux epoch for at least
one radio frequency. Only two sources have radio afterglows peaking at $>$50\,days at
$\sim$5\,GHz or higher, with the vast majority peaking at $<$10\,days. Typical sources
show a fall in flux of more than an order of magnitude between the first few days
and a rest frame epoch of 200\,days post-burst, with the flux evolution at 8.5\,GHz
being consistent with $f(t)\propto t^{-1}$ for times after peak flux.

Our late time observations of GRB\,100621A are entirely consistent
with those measured almost 200 days earlier in the galaxy rest
frame. The 2010 July observation remains discrepant but we note that
this was also the noisiest observation, and given the large errors,
the results are still consistent within 2\,$\sigma$ of our measured
value. Since the initial claim of a fading radio afterglow hinged
entirely on this relatively noisy datapoint, our late time observation
suggests that such an interpretation should be reconsidered.

 We cannot rule out an afterglow hypothesis, and further observations
 would be required to do so. However, given the lack of evidence for
 fading or other significant time variation in this source at either
 5.5 or 9.0\,GHz, we suggest that the constant flux is at least as
 likely to arise from underlying star formation in the host galaxy,
 rather than the burst afterglow. We make this assumption in the
 analysis that follows, while noting that it is inevitably subject to
 debate. Combining all available ATCA data in a single image (and
 thereby improving $uv$-plane coverage and sensitivity), we determine
 a best 5.5\,GHz flux for the host of GRB\,100621A of
 142$\pm$19\,$\mu$Jy, corresponding to a star formation rate of
 119$\pm$16\,M$_\odot$ yr$^{-1}$ (well within the range of star
 formation rates seen for the GRB population, see next section) -
 consistent with each of the reported individual measurements to
 within their errors. The source appears to be a point source at the
 resolution of the $3.2\times1.5''$ synthesised beam.

\begin{table}
\begin{center}
\begin{tabular}{lcccccccc}
Date &  Time post burst &  5.5\,GHz flux &  9.0\,GHz flux  \\ 
\hline\hline 
2010 Jun 24-25 &  1.6 & 137 $\pm$ 17  & 150 $\pm$ 28 \\
2010 Jun 25-26 &  2.3 & 129 $\pm$ 24  & 127 $\pm$ 45 \\
2010 Jul 17    &   17 & -43 $\pm$ 85  &  49 $\pm$ 100\\
2011 Apr 18    &  195 & 120 $\pm$ 32  & 106 $\pm$ 42 \\
\end{tabular}
\end{center}
\caption{ATCA observations of GRB\,100621A including those presented
  here for the first time, and those reported by
  \citet{2013A&A...560A..70G}. Fluxes and 1\,$\sigma$ flux errors
  (measured for a point source at the burst location) are given in
  $\mu$Jy/beam. Time after the {\em Swift} GRB trigger is given in
  days, calculated in the source {\em rest frame} (observed frame /
  1.5). \label{tab:100621a} }
\end{table}

\begin{figure}
\includegraphics[width=1.02\columnwidth]{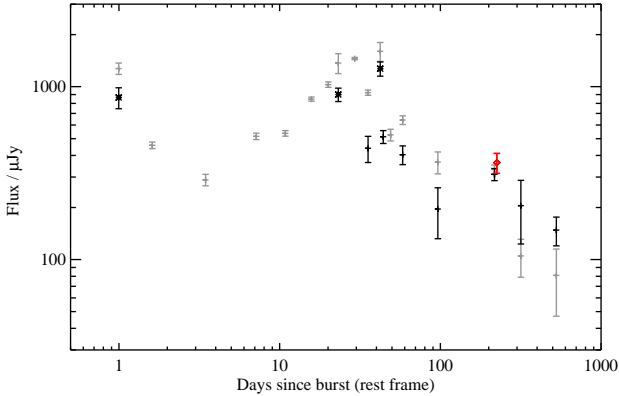}
\caption{ATCA and VLA observations of GRB\,100418A placing the observation
  reported here in the context of the radio afterglow light-curve reported by
  \citet{2013ApJ...779..105M}. Fluxes and 1\,$\sigma$ flux errors
  (measured for a point source at the burst location) are given in
  $\mu$Jy/beam. Observations at $\sim$5.5\,GHz are shown in black
  (asterisks: 5.5\,GHz from ATCA, crosses: 4.9\,GHz from VLA). Data at
  higher frequency ($\sim7-9$\,GHz) is shown in grey.  The 5.5\,GHz flux reported here
  is shown in red. Time after the {\em Swift} GRB trigger is given in the
  source rest frame. \label{fig:100418A}}
\end{figure}

The radio afterglow behaviour of GRB\,100418A has also been subject to
previous investigation, with a campaign of long term follow-up
conducted by \citep{2013ApJ...779..105M}. In figure \ref{fig:100418A},
we place our observation in the context of that study.  Interestingly,
our independent observation, taken with ATCA, confirms the relatively
high flux measured at the VLA twelve days before (in the observer
frame). This might suggest a puzzling late time rebrightening of this source
at $\sim$200 days post burst trigger in the rest frame.  Whatever the
implication of this high flux measurement, it seems clear that the
flux in this system is unlikely to be dominated by emission from the
GRB host galaxy and so cannot be interpreted as arising from star
formation.

\section{Star Formation Rates}\label{sec:SFRs}

The radio continuum flux from a galaxy can be converted to a star
formation rate, assuming that the emission is dominated by the
synchrotron component, which is emitted by electrons accelerated in
supernovae and their remnants.  Given that these events occur at the
end of the life of massive stars, the establishment of radio continuum
emission shows a delay relative to the ultraviolet emission from young
starbursts, but is directly related to star formation in older
populations with ongoing star formation.

Star formation rates determined for GRB host galaxies based on
optical-near-infrared emission are typically relatively low, of order
1-10\,M$_\odot$\,yr$^{-1}$ and are derived either from fitting to the
spectral energy distribution or to some assumed conversion factor from
ultraviolet continuum or H$\alpha$ line emission
\citep[e.g.][]{2009ApJ...691..182S,2010AJ....140.1557L,2011A&A...534A.108K}.
However the inferred star formation rate from fitting to the
UV-optical provides information primarily about the often blue,
relatively young star forming population dominating the emission at
these wavelengths, and does not provide any useful constraint on the
fraction of star formation taking place in obscured regions. A
relevant example is that of GRB 100621A, which has a measured
obscuration from the burst itself of A$_v=3.8$, while a fit to the SED
of the host yields just A$_V=0.6$ \citep{2011A&A...534A.108K}. The
star formation associated with the burst clearly occured in, or
behind, a heavily obscured region, and the optical data are not
representative of this extinguished population.

Since radio wavelengths are less affected by dust extinction than the
ultraviolet and optical, we might expect this flux to be a more
complete measurement of the galaxy star formation rate, and to exceed
the optically-derived values if substantial quantities of dust are
present in the host galaxy.

We convert our measured 5.5\,GHz fluxes to star formation rates, using the
known redshift of the sources, and the conversion prescription of \citet{2003ApJ...588...99B} and
\citet{2002ApJ...568...88Y} which depends on observed frequency, source redshift, and radio
spectral slope, $\alpha$.

Following \citeauthor{2003ApJ...588...99B}, we set $\alpha=-0.6$,
appropriate for faint radio sources. Using a steeper spectral slope,
$\alpha=-0.75$, results in star formation rate estimates 25\% higher
but does not significantly affect our conclusions. We fix the dust
temperature and emissivity index at $T_d=58$\,K and $\beta=1.35$
respectively (again for comparison with previous studies). The derived
star formation rates (in M$_\odot$\,yr$^{-1}$), are relatively
insensitive to these dust parameters for observations at $\sim$1-10\,GHz.

We over-plot contours on figure \ref{fig:radio_results} to show the
star formation rates equivalent to the given 5.5\,GHz flux as a
function of redshift. The last column in table \ref{tab:atca_obs}
gives the inferred star formation rate (or limit thereon) in solar
masses per year for each object, based on our 5.5\,GHz observations,
assuming that the observed radio flux is attributable to star
formation. The measurement of GRB\,100418A is not interpreted as star
formation, as discussed in section \ref{sec:afterglow}.

Given the very different beam shapes and sensitivities of the ATCA data, we do not attempt to 
stack undetected sources. However we do stack our three undetected sources with VLA data at
$z\sim1.1$. No source is detected in our stack, which has an rms flux of 1.9\,$\mu$Jy, 
 limiting the mean star formation rate of these three galaxies to $<16$M$_\odot$ yr$^{-1}$ (2\,$\sigma$).

We note that, while the host of GRB\,050223 is detected at 3.4 and 4.6
$\mu$m, none of the objects in our sample are detected in 22$\mu$m
$W4$-band infrared data from the ALLWISE data release
\citep{2010AJ....140.1868W,2011ApJ...731...53M}. Flux in the long
wavelength WISE bands is also excited by star formation and can be
used as a star formation rate indicator, although this relies on
calibrations that have been determined locally, and will vary as a
function of redshift.  At the redshift of our four detected sources,
the $22\,\mu$m is probing the rest frame region around 14\,$\mu$m -
and so is sensitive to hot dust and PAH emission.  While strong flux
in this region is indicative of a strong photon source - often star
formation - the conversion is dependent on the physical properties of
the star formation region and is thus poorly calibrated. WISE is also
a relatively shallow survey, sufficient to identify very luminous
sources, but not typical star forming galaxies at $z>0.5$.  In table
\ref{tab:wise}, we give flux limits and equivalent star formation rate
estimates for the four sources with proposed radio detections in this
work. We apply the star formation rate conversion factor calculated
from $z=0-0.3$ galaxies \citep{2013ApJ...774...62L} for luminosities
measured in the broad $W3$ filter, which spans 11-16$\mu$m (observed)
and thus provides a better calibration for our distant sources than
the low redshift $W4$ band. As can be seen, the limits obtained by
WISE are consistent with our inferred radio star formation rates for
these sources, and also consistent with none of the larger sample
being heavily obscured, submillimeter galaxy-like intense starbursts.

\begin{table}
\begin{center}
\begin{tabular}{lcccccccc}
Target  & $\lambda_\mathrm{rest}$& Flux &  SFR  \\ 
 & $\mu$m & mJy &  M$_\odot$\,yr$^{-1}$ \\
\hline\hline 
GRB\,050223  & 13.8 & $<$ 1.9   &  $<$ 320 \\ 
GRB\,060729  & 14.3 & $<$ 2.2   &  $<$ 290 \\ 
GRB\,100418A & 13.6 & $<$ 1.2   &  $<$ 220 \\ 
GRB\,100621A & 14.3 & $<$ 2.4   &  $<$ 320 \\ 
\end{tabular}
\end{center}
\caption{WISE limits on observed 22$\mu$m ($W4$) flux and inferred star formation rate for the four sources
 with possible radio detections. 2$\sigma$ limits are given in each case. \label{tab:wise} }
\end{table}

An additional constraint exists for one of our detected sources - the
host of GRB\,050223 - in the form of mid-infrared {\em Herschel Space
  Telescope} data. \citet{2014arXiv1402.4006H} observed a small sample
of `dark' GRBs with the PACS and SPIRE instruments, obtaining
constraints on their mid-infrared spectral energy distribution (SED).
In the case of GRB\,050223, the host galaxy remained undetected, and
the SED fit is driven by the optical to near-infrared data observed
through to the WISE $W1$ and $W2$ bands. As figure 2 of
\citet{2014arXiv1402.4006H} makes clear, the Herschel data do not
tightly constrain the SED of this target, and derived parameters such
as dust temperature and emission spectrum, star burst fraction, star
formation rate and obscuration are highly degenerate. In the case of
GRB\,050223 the fit to the stellar spectrum dominated optical does
little to break these degeneracies or constrain the longer wavelength
emission. Indeed the infrared star formation rate determined by these
authors is somewhat puzzling, actually falling below that estimated
from the (uncorrected) ultraviolet emission. Given that this host
galaxy had already been identified as a dusty starburst, with an
A$_V>2$ \citep[i.e. extinguished by a factor $>5$ in the
  ultraviolet,][]{2006A&A...459L...5P}, one might expect an estimate
of the infrared emission to exceed the ultraviolet estimate by this
factor, even without accounting for any more obscured emission. Thus
while noting that our estimate of star formation significantly exceeds
that of \citeauthor{2014arXiv1402.4006H}, we suggest that this
indicates that further investigation of this source, or a revision to
the existing spectral energy distribution fits, might be appropriate.

\section{Discussion}\label{sec:discussion}

 \subsection{Redshift Evolution}\label{sec:z}

\begin{figure*}
\includegraphics[width=1.8\columnwidth]{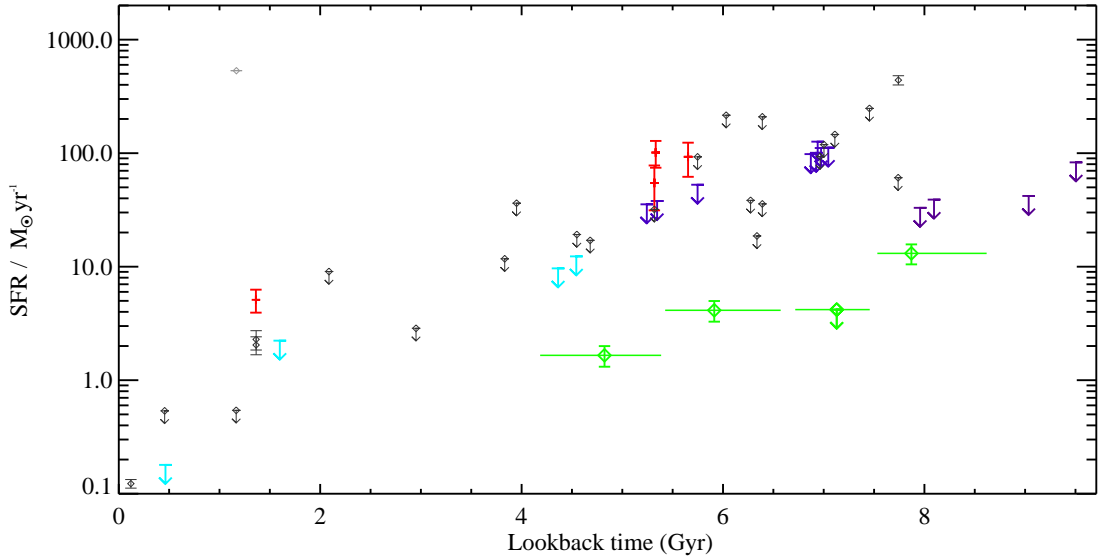}
\caption{The radio-inferred star formation rates (and limits) for GRB
  host galaxies with radio observations, shown as a function of
  lookback time. Large symbols indicate 5.5\,GHz measurements taken as part
  of this programme from \citet{2010MNRAS.409L..74S} at $z<0.5$ (lookback $<$5\,Gyr), and
  from the observations presented in this paper from ATCA
  ($0.5<z<1.0$, 5-7.5\,Gyr) and the VLA ($z>1$, $>$7.5\,Gyr). Small symbols indicate the
  measurements reported or compiled by \citet{2012ApJ...755...85M} -
  triangles mark measurements at $<5$\,GHz and diamonds those at
  higher frequency. Note that several sources have measurements at
  multiple frequencies, so not every point is independent. All limits are
  at the 2\,$\sigma$ level. Large green
  diamonds indicate the average rates inferred for the hosts of core collapse
  SNe in the GOODS fields as described in section
  \ref{sec:ccsn}, with horizontal bars showing the limits of each redshift
  bin. GRB\,060814 lies at $z=1.92$ and is omitted from this figure.\label{fig:zevol}}
\end{figure*}

 In figure \ref{fig:zevol} we consider the GRB host star formation
 rate measurements and limits derived from radio emission, as a
 function of lookback time.  In addition to the data reported in this paper
 and our previous work \citep{2010MNRAS.409L..74S}, we also plot the
 radio-derived star formation rates reported and compiled by
 \citet{2012ApJ...755...85M}. To enable direct comparison, we adjust
 the star formation rates of \citet{2012ApJ...755...85M} to account
 for the steeper radio spectral slope assumed in that paper, and to
 report 2\,$\sigma$ limits where a source is undetected.

 In total, figure \ref{fig:zevol} includes data on 40 long gamma ray
 burst host galaxies. Of these, a mere six have secure detections
 (excluding the afterglow of GRB\,100418A), and two of those lie at
 $z<0.15$ (i.e. 2\,Gyr lookback time), making clear the difficulty of
 observing these low mass, moderately star forming galaxies over
 substantial intervals of cosmic time.

  Five sources have only modest constraints (a 2\,$\sigma$ limit
  on star formation rate $>$200\,M$_\odot$ yr$^{-1}$) and a sixth
  source is detected with a very high star formation rate
  ($\sim$400\,M$_\odot$ yr$^{-1}$). This source, GRB\,021211 is a
  pre-{\em Swift} burst, arguably with different selection criteria
  and characteristics to the bulk of the sample. We note that the high
  flux density for this source (S$_{1.4GHz}=330\pm31$\,$\mu$Jy) reported
  by \citet{2012ApJ...755...85M} is inconsistent with that reported by
  \citep{2012ApJ...748..108H}, who constrain the star formation rate
  to $\la60$\,M$_\odot$ yr$^{-1}$ (2\,$\sigma$, with our
  assumptions). Further investigation of this source is clearly
  needed.

  In summary, no more than 5 of the 40 ($12\pm6$\%) GRB hosts with
  radio constraints, have star formation rates exceeding
  200\,M$_\odot$ yr$^{-1}$. This confirms, and indeed strengthens, the
  findings of \citet{2012ApJ...755...85M} who analysed data on about
  half of this composite sample, and is somewhat lower than
  suggested by early, pre-{\em Swift} and higher redshift work by
  \citet[who suggested a fraction of 20\%]{2003ApJ...588...99B} and
  \citet[who detected 3 of 21 sources]{2004MNRAS.352.1073T}. The
  reasons for this remain unclear, although small number statistics
  and the selection criteria for follow-up targets no doubt contribute
  to the discrepency.

  The detection fraction in our $z\sim0.5$ (lookback time
  $\sim$5.5\,Gyr) ATCA observations is perhaps slightly higher than
  might be expected based on the the statistics of the sample as a
  whole.  Of nine GRB hosts at $0.45<z<0.70$ with star formation rate
  limits better than 200\,M$_\odot$\,yr$^{-1}$ we detect three sources
  (33$\pm$19\%). While these statistics are still based on small
  numbers of objects, if they were typical of the sample as a whole we
  might have expected several more detections outside this redshift
  range, including more examples at $z<0.5$ and perhaps one of our
  $z\sim1$ targets.

  At present, the number counts in this subsample are insufficient to
  extrapolate further, and we note that one of these is our least
  significant detection. However, it suggests that further
  investigation may be warranted to determine whether this anomaly
  resolves with added data, or remains statistically significant. If
  there is, in fact, a redshift dependence in the typical star
  formation rate of GRB host galaxies, then it is possible that
  $z\sim0.5$ may represent a sweet spot in current telescope sensitivity
  relative to the typical star formation rate in GRB host galaxies.
  At lower redshifts, GRBs may be occuring in galaxies with lower typical
  star formation rates, while at higher redshift, the current radio limits
  are too weak to reliably probe this regime.
  
  We note that other than this anomaly, there is at best very modest
  evidence for redshift evolution in the radio derived star formation
  rates in individual GRB host galaxies. The measurement sample is
  dominated by upper limits at all redshifts, but these are less
  constraining with increasing redshift. Nonetheless the upper limits
  on typical sources at lookback times of 6-10\,Gyrs (roughly
  corresponding to $z=0.65-1.85$) constrain the typical GRB host
  radio-derived star formation rate to be no more than an order of
  magnitude higher than that observed at $z<0.15$. This is comparable
  to the expected change in the typical star formation rate for
  galaxies of a given mass over the same interval due to the effects
  of downsizing \citep[see, for example,][and references
    therein]{2014arXiv1405.2041S}, and so while we limit the redshift
  evolution of GRB hosts, the current sample is insufficient to
  conclude that their evolution is significantly different to that of
  typical star forming galaxies over cosmic time.

 \subsection{Core Collapse SN Hosts}\label{sec:ccsn}  

 Long GRBs are believed to be generated by the breakout of
 relativistic jets during the collapse of of a massive star at the end
 of its life \citep{2006ApJ...637..914W}.  They have been observed to be coincident with Type Ic
 supernovae, and may be associated with core collapse supernovae
 (CCSN) more generally  \citep[see][and references therein]{2006Natur.441..463F}.

 However, as discussed in section \ref{sec:intro}, \citet{2006Natur.441..463F} found that GRB hosts are
 typically less optically luminous than those of CCSNe, and also that the
 transient is more closely
 associated with the peak of the light distribution in the host, based
 on optical photometry with the {\em Hubble Space Telescope}. They
 suggested that this might well represent a difference in the
 metallicity of the progenitor population, with GRBs biased towards
 the most massive and lowest metallicity stellar populations. This
 conclusion was broadly supported by \citet{2010MNRAS.tmp..479S}, who
 employed template fitting of the spectral energy distribution of the
 host galaxies to make the same comparison. However
 \citet{2010MNRAS.tmp..479S} concluded that the difference between the
 two host populations was less pronounced in the blue optical bands
 which are dominated by star formation. This suggests the intriguing
 possibility that in the star formation-dominated radio continuum, the
 two populations may again appear similar.

 We consider the published catalogue of core-collapse SNe in the GOODS
 field
 \citep{2004ApJ...613..200S,2008ApJ...681..462D,2012ApJ...757...70D},
 matching that also analysed by \citet{2010MNRAS.tmp..479S} and
 omitting only sources without radio coverage. These sources were
 initially selected, based on detection of the optical supernova, to
 lie in the two GOODS survey fields due to their extensive
 multi-wavelength coverage with {\em HST}
 \citep{2004ApJ...600L..93G}. Both GOODS fields have also been
 surveyed with deep radio imaging at 1.4\,GHz from the VLA. We make
 use of the publically released maps of GOODS-S from
 \citet{2008ApJS..179..114M} and GOODS-N from
 \citet{2010ApJS..188..178M}. Both images have a comparable
 sensitivity, and are on the same pixel scale. While several core
 collapse SN hosts show a marginal detection, no individual source is
 strongly detected.  We combine the radio flux at the location of
 those CCSN in the sample that lie within the appropriate image to
 produce stacked images, taking the mean pixel value at each point.

\begin{figure}
\includegraphics[width=0.99\columnwidth]{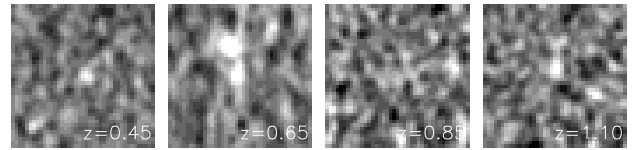}
\caption{The stacked radio flux at 1.4\,GHz for core collapse SN hosts
  in the GOODS fields, as described in section \ref{sec:ccsn}. The
  average CCSN host is well detected in three redshift bins. The third
  bin, centred at $z\sim0.85$ does not show any evidence for a
  detection and is treated as an upper limit. The bright off-centre
  source in the $z\sim0.65$ stack is the result of bright neighbours
  in two individual images biasing the mean. \label{fig:ccsn_stack}}
\end{figure}

 We divide the sample of CCSN in \citet{2010MNRAS.tmp..479S} into four
 redshift bins, each with $\Delta z=0.2$, and combine the objects in
 each separately.  The resulting stacked images are shown in figure
 \ref{fig:ccsn_stack} and their measured properties given in table
 \ref{tab:ccsn}. Radio properties of individual GOODS supernova hosts
 are given in the appendix and not discussed further here. In three
 redshift bins the average CCSN host galaxy is well detected. The
 $z\sim0.85$ sample has no clear detection, and the measured flux is
 treated as an upper limit.

 The GOODS-S field was observed by the VLA in Jun-Sep 2007,
 $>$2\,years after the last supernova was detected and host galaxy
 measurements are unlikely to be contaminated by residual flux from the supernovae.  The
 GOODS-N observations represent a combination of data from 1996 and
 Feb 2005-Apr 2006. We cannot rule out the contribution of supernova
 flux to these observations, but note that $\sim78$\% of the GOODS-N
 data were collected either before the SNe were detected, or
 $>1$\,year post-SN. The radio lightcurves of core-collapse supernovae
 decay on comparable timescales to those of GRBs but are typically at
 least two orders of magnitude less luminous than GRBs at comparable
 redshift \citep{2014arXiv1401.1221K}. Given the typical peak radio
 luminosity of $2\times10^{27}$ ergs\,s$^{-1}$\,Hz$^{-1}$ and a time
 decay of index $\sim-1$ \citep{2002ARA&A..40..387W}, we would expect
 radio supernovae to have dropped well below our stacked detection
 limit within a year in the observed frame, at all redshifts
 considered here.  Assuming then that the host galaxy flux dominates
 over any supernova contribution, we convert the fluxes to star
 formation rates, using the same prescription given earlier to account
 for the different observed-frame frequency of the observations
 (1.4\,GHz rather than 5.5\,GHz).

\begin{table}
\begin{center}
\begin{tabular}{cccc}
$<z>$ (Range)         & Number &  Flux$_\mathrm{1.4\,GHz}$ / $\mu$Jy & SFR / M$_\odot$\,yr$^{-1}$ \\
\hline\hline
0.47 (0.35-0.55)      & 17 & 6.0 $\pm$ 1.2 & 1.7  $\pm$   0.3 \\
0.63 (0.55-0.75)      & 17 & 7.6 $\pm$ 1.6 & 4.1  $\pm$   0.9 \\
0.86 (0.75-0.95)      &  8 & 3.8 $\pm$ 1.9 &  $<$ 4.2         \\
1.04 (0.95-1.15)      &  9 & 7.9 $\pm$ 1.6 & 13.1   $\pm$   2.6 \\
\end{tabular}
\end{center}
\caption{The number of sources and measured radio properties of the CCSN host galaxy stacks 
discussed in section \ref{sec:ccsn}. \label{tab:ccsn}}
\end{table}

 The derived star formation rates are compared to those of the GRB
 host sample and shown as large green diamonds in figure
 \ref{fig:zevol}. The CCSNe typically occur in galaxies with lower star formation
 rates than are reached by current samples, confirming that the
 scarcity of detections in our sample is unsurprising. Interestingly,
 however, the results of the core collapse supernovae host sample are
 consistent with an increase in the typical radio-derived star
 formation rate in these galaxies with increasing redshift.

  This behaviour appears consistent with the expected picture of cosmic
  `downsizing' \citep{1996AJ....112..839C}. Not only does the volume-averaged
  star formation density of the Universe increase by more than an order of magnitude
  with increasing lookback time from $z=0$ to $z=1$ but the size and star
  formation rate of galaxies in the star-forming population also increases. Thus, if
  core-collapse supernova rate is directly proportional to star formation rate (i.e. the
  larger core collapse population is an unbiased tracer, as GRBs are not believed to be), the
  majority of events would be expected to be seen in more intense starbursts at $z\sim1$
  than typically exist in the local Universe.
  As a result the detection of a single core-collapse supernova implies the existence of
  more young stars if it occurs at $z\sim1$ than if it occurs more locally, and any attempt
  to reconstruct the cosmic star formation history would require an adjustment for this evolution
  in the host galaxy population.

 If GRBs are indeed still more biased towards low metallicity
 environments, as theoretically and observationally suggested
 \citep[e.g.][]{2006ApJ...637..914W,2010AJ....140.1557L,2013ApJ...774..119G},
 but not universally accepted \citep[e.g.][]{2010ApJ...712L..26L},
 then this effect would likely be stronger, with lower star formation
 rates needed, on average, to produce a burst at high redshift than in
 the local Universe, since a fraction of the higher metallicity star formation
is not represented in the GRB rate. Further observations are clearly required to
 determine whether GRB hosts follow a similar pattern to that seen in
 the small sample of CCSN presented here, which could plausibly bias
 attempts to interpret the burst rate at a given redshift as a proxy
 for the volume averaged cosmic star formation rate. We note that the increase
in radio-derived star formation rate between our lowest redshift bin and the highest
is a factor of 7.7$^{+3.5}_{-2.5}$.  If the typical rates seen in GRBs in the local
Universe increase by a similar factor, it would be doubtful whether our radio observations
are sufficiently sensitive to detect them.

 In a pessimistic scenario, where GRB hosts are no more radio luminous than those of the 
 larger CCSN population at high redshift, securing radio detections at the level of the core
 collapse supernovae hosts would require an improvement in sensitivity
 by a factor of 10 over the current observations. While using a larger
 array (for example the VLA rather than ATCA) may allow progress to be
 made in reasonable integration times at $z\sim0.5$, the hosts at
 $z\sim1$ will need integration times approaching 8-10 hours per source to 
 reach star formation rates of $\sim$10\,M$_\odot$\,yr$^{-1}$,
 making this an expensive study to undertake.

 \subsection{Dark GRB Hosts}\label{sec:dark}

 Four of our sample, GRBs\,051022, 060814, 050223 and 100621A, have
 been identified as dark bursts in earlier work. These are defined as
 sources which are sub-luminous in the optical relative to the X-ray,
 typically with an optical-to-X-ray spectral slope
 $\beta_{OX}<0.5$. This spectral index is beyond those
normally allowed in GRB fireball models, and is most commonly 
attributed to the presence of dust in the GRB host galaxy. Indeed, 
the majority of these dark bursts 
have been shown to be associated with high optical extinctions
\citep{2013ApJ...778..128P}. Studies in the
 optical and near-infared have suggested that the hosts of such bursts
 are typically more massive and chemically evolved than those of other
 GRBs, with higher star formation
 rates \citep[e.g.][]{2013ApJ...778..128P,2011A&A...534A.108K}.
 However, as already discussed, dust obscuration can make
 intepretation of galaxy spectral energy distributions
 challenging. Given the difficulties associated with faint optical
 emission, radio continuum observations have the potential to improve
 our understanding of the host galaxies of these
 sources \citep{2013ApJ...767..161Z}.

Two of our candidates, GRBs\,051022 and 060814 are included in the
dark burst sample of \citet{2013ApJ...778..128P}. Another,
GRB\,050223, does not satisfy the strict criteria for $\beta_{OX}$,
with the strongest available constraint of $\beta_{OX}<0.8$, but based
on its host galaxy properties \citet{2006A&A...459L...5P} claim this
source as part of the dark sample \citep[although][find no evidence
for excess extinction in the X-ray emission from the
burst]{2005MNRAS.363L..76P}. A high extinction ($A_V$=3.8) and low
$\beta_{OX}=0.278$ has also been reported for a fourth source,
GRB\,100621A \citep{2011A&A...534A.108K,2012MNRAS.421.1265M}.

 Radio constraints on the host galaxies of an exclusively `dark'
 sample of 15 hosts, observed from the VLA, have recently been
 presented by \citet{2013ApJ...778..172P}. Only one source,
 GRB\,051022, overlaps with the sample presented here, with the
 majority of targets in the Perley \& Perley study lying at higher
 redshift than those in our sample.  Four of their targets were
 detected: 3 host galaxies at SFR$>800$\,M$_\odot$\,yr$^{-1}$ and
 GRB\,051022 at $74\pm20$\,M$_\odot$\,yr$^{-1}$, consistent with our
 limit given in table \ref{tab:atca_obs}. Two further host galaxies
 (GRBs\,060202 and 090417B) were constrained to have radio star
 formation rates under 120\,M$_\odot$\,yr$^{-1}$, while the remaining,
 higher redshift targets, have less constraining upper limits on their
 star formation rates spanning from $<325$ to a relatively weak
 $<4140$\,M$_\odot$\,yr$^{-1}$.

 We detect the hosts of dark GRBs\,050223 and 100621A in our 5.5\,GHz
 observations, giving radio-derived star formation rates of
 $\sim90-100$\,M$_\odot$\,yr$^{-1}$ -- significantly higher than the
 majority of GRB hosts. Given the likely high redshift of GRB\,060814
 (see section \ref{sec:sample}), the star formation rate constraint on
 this source is rather weak and we exclude it from further discussion.
 
Between the \citet{2013ApJ...778..172P} study and the observations
reported here, the hosts of nine dark bursts have detected or well
constrained radio derived SFRs - three in the range SFR$\sim70
100$\,M$_\odot$\,yr$^{-1}$, three with higher star formation rates
more akin to ULIRGs ($>200$\,M$_\odot$\,yr$^{-1}$), and three
undetected sources constrained to have SFR$<120$\,M$_\odot$\,yr$^{-1}$
(typical of the GRB host population as a whole, see
figure \ref{fig:zevol}). The remainder (lying at higher redshift)
have relatively weak constraints on their radio flux and
could still be harbouring star formation rates of several hundred
solar masses per year, and yet remain undetected. As with
GRBs\,060814, we exclude these from further
analysis.

 To evaluate the probability of this detection rate occuring by
 chance, assuming that dark burst hosts are drawn from the same
 underlying star formation rate distribution as the non-dark burst
 sample, we consider a bootstrap resampling of the full existing,
 non-dark burst, dataset.  The observations presented here and
 in \citet{2012ApJ...755...85M} provide detections or limits of
 $<120$\,M$_\odot$\,yr$^{-1}$ or better on 28 individual non-dark
 sources. We make the conservative assumption that we have barely
 missed detecting the majority of sources (i.e. that their 2\,$\sigma$
 limits are representative of their true star formation rates), and
 randomly draw sub-samples of nine objects from this compilation to
 mimic the dark burst sample, repeating the process $10^7$ times. We
 find that in only 1.7\% of cases do six or more randomly sampled
 objects in a sample of nine have star formation rates exceeding
 60\,M$_\odot$\,yr$^{-1}$. As such, it is highly unlikely that the
 dark busts we have identified above, are drawn from same underlying
 population as the non-dark bursts. The deviation of dark burst host
 star formation rates from those of the non-dark GRBs, is significant
 at the 2\,$\sigma$ level (subject, of course, to the vagaries of low
 number statistics) and merits further investigation.

This may suggest that while, as \citet{2013ApJ...778..172P} and others
 have concluded, dark GRB hosts are typically forming stars at a lower
 rate than submillimeter galaxies, they are nonetheless biased towards
 systems with higher star formation rates than seen in the general,
 less extinguished long GRB host population.

\section{Conclusions}\label{sec:conc}

Our conclusions can be summarized as follows:

\begin{enumerate}
\item We present observations of 17 gamma ray burst host galaxies at 5.5\,GHz, with 13 of these also observed
 simultaneously at 9.0\,GHz. Typical rms noise levels in the images are $\sim$30\,$\mu$Jy/beam.

\item Radio continuum (5.5\,GHz) point sources are detected at the location of  four GRBs (GRB\,050223, 060729, 100418A and 100621A). Assuming a standard conversion factors, the fluxes of these correspond to radio-derived star formation rates of $\sim60-200$\,M$_\odot$\,yr$^{-1}$.

\item Comparison of our late time measurements with published data suggests that the radio flux reported at early times by \citet{2013A&A...560A..70G} in GRB\,100621A likely arises from the host galaxy rather than the burst afterglow as previously suggested. However our detection of GRB\,100418A likely arises from late time radio afterglow emission. Our remaining two sources are observed several years post-burst.

\item Based on our observations, we see no strong evidence for evolution in the typical star formation rate of the GRB host galaxy population with redshift, but note that we detect three sources at $z\sim0.5$ - an anomalously high fraction of sources in that redshift bin. Given the small number statistics it is difficult to comment further on the significance of this. 

\item We compare to the typical radio emission of CCSN hosts from the GOODS survey, securing detections in stacked samples at $z\sim0.47, 0.63$ and 1.04, but not at $z\sim0.86$. These detections correspond to star formation rates that are substantially lower (by an order of magnitude) than current GRB host limits. They also show some evidence for a trend towards higher star formation rates at higher redshift.

\item Four of our targets satisfy criteria for identification as dark bursts. We detect two of these. In combination with earlier results, we suggest that these sources may have average star formation rates rather higher than those seen in the general population of long GRB hosts.

\end{enumerate}

\section*{Acknowledgments}
This paper is based in part on data obtained at the Australia Telescope Compact Array associated with programme C2544. The Australia Telescope Compact Array is part of the Australia Telescope National Facility which is funded by the Commonwealth of Australia for operation as a National Facility managed by CSIRO.

Also based in part on observations taken at the NRAO Karl G. Jansky Very Large Array (VLA), associated with programme 12A-279. The National Radio Astronomy Observatory is a facility of the National Science Foundation operated under cooperative agreement by Associated Universities, Inc.

This work made use of data supplied by the UK Swift Science Data Centre at the University of Leicester \citep{2009MNRAS.397.1177E}. We also made use of Ned Wright's very useful cosmology calculator \citep{2006PASP..118.1711W}.

AJL and ERS acknowledge support from the UK Science and Technology Facilities Council, under the Warwick Astrophysics consolidated grant ST/L000733/1 and PATT-linked travel support grant. We also acknowledge the influence of suggestions from the anonymous referee on the evolution of this paper.

\bsp

\appendix

\section{The hosts of Core Collapse Supernovae}\label{sec:appendix}

As discussed in section \ref{sec:ccsn}, we extract radio fluxes at the
locations of core-collapse supernovae originally identified as part of
the GOODS campaign \citep{2004ApJ...600L..93G}.  Sources identified
during the 2002-2003 observing campaign were catalogued, classified
and reported by \citet{2004ApJ...613..200S}, listed both by `nickname'
and IAU approved transient name.  Sources from 2004-2005 were classified by and
discussed in \citet{2008ApJ...681..462D} and
\citet{2012ApJ...757...70D}.  Formal IAU naming was not sought for
these later objects. Spectroscopic redshifts were obtained either from
literature values for the supernova host galaxy or through Keck
Observatory spectroscopy primarily originating from or compiled by the
Team Keck Treasury Redshift Survey \citep[][]{2004AJ....127.3121W}. Target classifications, locations and
redshifts for all these sources were made available electronically in
2008, as described in \citet{2008ApJ...681..462D}, and we make use of
this publically released catalogue, and specifically those supernovae 
classified as core collapse and within the coverage region of the 
available radio imaging.

In this appendix we 
tabulate the measured radio fluxes at each position 
in table \ref{tab:sne_data}. Radio fluxes are measured on the
publically released 1.4\,GHz VLA maps of GOODS-S from
\citet{2008ApJS..179..114M} and GOODS-N from
\citet{2010ApJS..188..178M}.
Names presented are those used
in previous literature. Where formal names are available these are given, else the GOODS team `nickname' is presented for ease of comparison with previous work. For details of the sample see
\citet{2012ApJ...757...70D}, see also \citet{2010MNRAS.tmp..479S} for
further analysis of these sources. Note
that only objects with redshifts between $z=0.35$ and $z=1.35$
contribute to the stacks described in section \ref{sec:ccsn}.
Six supernova host galaxies are individually detected at better than
3\,$\sigma$, with the strongest two detections both lying at the low
redshift end of the sample. The remaining objects do not represent
individual detections at the supernova location.

\begin{table*}

\begin{tabular}{llrrrrr}
\hline
  ID & RA \& Declination (J2000) & Redshift & Flux & RMS & S/N &	\\
 & & & $\mu$Jy & $\mu$Jy\\
\hline
  2002fz 	&  03 32 48.566	-27 54 17.73	    & 0.838 	& 6.9 	& 7.9 	& 0.9	\\
  2002hq 	&  03 32 30.027	-27 43 47.40	    & 0.669 	& 10.3 	& 9.1 	& 1.1	\\
  2002hs 	&  03 32 18.537	-27 48 34.14	    & 0.388 	& -3.8 	& 7.6 	& -0.5	\\
  2002ke 	&  03 31 58.677	-27 45 00.32	    & 0.577 	& 8.5 	& 7.6 	& 1.1	\\
  2002kb 	&  03 32 42.429	-27 50 25.15	    & 0.578 	& 26.9 	& 6.8 	& 3.9&$\ast$	\\
  2002kl 	&  12 37 49.281	 62 14 06.61	    & 0.412 	& 5.6 	& 6.0 	& 0.9	\\
  2003ba 	&  12 36 15.912	 62 12 37.70	    & 0.286 	& 9.7  	& 3.9 	& 2.5	\\
  2003bb 	&  12 36 24.423	 62 08 36.58	    & 0.955 	& 8.0 	& 4.4 	& 1.8	\\
  2003bc 	&  12 36 38.210	 62 09 53.78	    & 0.511 	& 4.7 	& 4.0 	& 1.2	\\
  2003dx 	&  12 36 31.681	 62 08 48.66	    & 0.512 	& 5.7 	& 5.0 	& 1.1	\\
  2003dz 	&  12 36 39.921	 62 07 52.56	    & 0.48 	        & 2.7 	& 3.8 	& 0.7	\\
  2003en 	&  12 36 33.149	 62 13 47.66	    & 0.54 	        & -0.81 & 4.0 	& -0.2	\\
  2003er 	&  12 36 32.384	 62 07 34.48	    & 0.595 	& 15.5 	& 4.1 	& 3.8&$\ast$	\\
  2003et 	&  12 35 55.862	 62 13 33.13	    & 1.296 	& 8.4 	& 3.7 	& 2.3	\\
  2003ew 	&  12 36 27.806	 62 11 25.07	    & 0.517 	& 5.1 	& 4.0 	& 1.3	\\
  2003N 	&  12 37 09.265	 62 11 00.65	    & 0.425 	& 3.0 	& 6.3 	& 0.5	\\
  K0404-005 	&  12 36 27.056	 62 15 09.75	    & 0.79 	        & -1.4 	& 3.8 	& -0.4	\\
  K0404-006 	&  12 37 06.731	 62 21 17.82	    & 0.406 	& 4.0 	& 4.4 	& 0.9	\\
  K0404-008 	&  12 38 03.647	 62 17 11.77	    & 0.278 	& 69.0  & 3.9 	& 17.5&$\ast$	\\
  K0404-010 	&  12 36 46.069	 62 16 25.80	    & 0.61 	        & -2.2 	& 5.3 	& -0.4	\\
  K0404-011 	&  12 36 49.362	 62 16 04.83	    & 1.53 	        & 1.9 	& 4.0 	& 0.5	\\
  K0405-001 	&  12 35 50.757	 62 10 37.95	    & 1.012 	& 5.6 	& 3.8 	& 1.5	\\
  K0405-002 	&  12 36 26.710	 62 08 30.15	    & 0.556 	& 0.40 	& 4.4 	& 0.1	\\
  K0405-005 	&  12 36 35.991	 62 17 32.67	    & 0.683 	& -0.12 & 4.1 	& -0.03	\\
  K0405-007 	&  12 36 58.447	 62 16 37.37	    & 0.497 	& -1.2 	& 4.0 	& -0.3	\\
  K0405-008 	&  12 37 33.842	 62 19 22.24	    & 0.88 	        & 2.3 	& 5.0   & 0.5	\\
  HST04Bon 	&  03 32 16.164	-27 49 41.81	    & 0.664 	& 34.0 	& 6.9 	& 4.9&$\ast$	\\
  HST04Cli 	&  03 32 05.073	-27 41 42.61	    & 0.75 	        & 6.82 	& 7.1 	& 1.0	\\
  HST04Con 	&  12 37 08.258	 62 12 53.74	    & 0.838 	& 6.1 	& 3.9 	& 1.6	\\
  HST04Cum 	&  12 37 18.453	 62 10 50.72	    & 0.972 	& 4.8 	& 3.8 	& 1.2	\\
  HST04Fox 	&  03 32 41.765	-27 53 29.45	    & 0.69 	        & 11.8 	& 6.8 	& 1.7	\\
  HST04Geo 	&  12 36 44.432	 62 10 53.19	    & 0.937 	& 2.2 	& 4.2 	& 0.5	\\
  HST04Gua 	&  12 37 36.423	 62 16 27.12	    & 1.26 	        & 3.3 	& 4.0 	& 0.8	\\
  HST04Hei 	&  03 32 42.415	-27 50 21.76	    & 0.576 	& 4.9 	& 7.0 	& 0.7	\\
  HST04Jef 	&  03 32 23.679	-27 53 27.67	    & 0.964 	& 10.0 	& 6.3 	& 1.6	\\
  HST04Ken 	&  03 32 44.702	-27 49 22.77	    & 0.522 	& 13.0 	& 6.0 	& 2.2	\\
  HST04Mur 	&  03 32 35.307	-27 49 39.95	    & 1.79 	        & 3.5 	& 6.5 	& 0.5	\\
  HST04Pata 	&  12 37 25.258	 62 10 06.36	    & 0.41 	        & 32.4 	& 4.1 	& 8.0&$\ast$	\\
  HST04Patu 	&  12 38 08.961	 62 18 47.39	    & 0.571 	& 4.5 	& 4.4 	& 1.0	\\
  HST04Pol 	&  03 32 19.681	-27 50 23.53	    & 0.561 	& 8.6 	& 6.8 	& 1.3	\\
  HST04Riv 	&  03 32 32.407	-27 44 52.84	    & 0.606 	& 8.6 	& 6.4 	& 1.3	\\
  HST04Sos 	&  03 32 22.638	-27 50 15.30	    & 0.55         	& 4.6 	& 5.5 	& 0.8	\\
  HST04Tov 	&  03 32 49.625	-27 55 34.73	    & 1.83 	        & 10.7 	& 6.7 	& 1.6	\\
  HST04Wil 	&  03 32 13.061	-27 42 04.92	    & 0.422 	& -6.9 	& 7.2 	& -0.9	\\
  HST05Boy 	&  03 32 44.859	-27 54 11.25	    & 0.66 	        & -10.6 & 7.3 	& -1.5	\\
  HST05Bra      &  12 37 21.764	 62 12 25.67	    & 0.48 	        & 10.4 	& 4.9 	& 2.2	\\
  HST05Cas 	&  12 36 07.767	 62 13 08.63	    & 0.73 	        & 2.43 	& 3.9 	& 0.6	\\
  HST05Den      &  12 37 14.773	 62 10 32.61	    & 0.971 	& 5.2 	& 4.3 	& 1.2	\\
  HST05Fil 	&  12 37 19.329	 62 15 59.30	    & 1.21 	        & 2.9 	& 4.0   & 0.7	\\
  HST05Kir 	&  12 36 14.875	 62 12 53.07	    & 0.448 	& 6.3 	& 3.9 	& 1.6	\\
  HST05Mob 	&  12 36 25.518	 62 15 11.04	    & 0.68 	        & 1.3 	& 3.6 	& 0.4	\\
  HST05Pic 	&  12 37 37.965	 62 12 43.95	    & 0.911 	& 4.0 	& 4.1 	& 1.0	\\
  HST05Sco 	&  12 36 51.196	 62 19 56.11	    & 0.93 	        & -7.6 	& 4.2 	& -1.8	\\
  HST05Sev 	&  12 37 43.773	 62 14 38.11	    & 0.96 	        & 4.6 	& 4.0 	& 1.2	\\
  HST05Ste 	&  12 37 01.534	 62 17 47.13	    & 0.475 	& 3.5 	& 4.5 	& 0.8	\\
  HST05Str      &  12 36 46.879	 62 11 45.21	    & 1.06 	        & 2.3 	& 4.3 	& 0.6	\\
  HST05Ton 	&  12 37 01.545	 62 11 29.01	    & 0.778 	& 16.4 	& 4.2 	& 3.9&$\ast$	\\
\hline\end{tabular}

\caption{Measured 1.4\,GHz data at the location of the core collapse
  supernovae discussed in this paper. Fluxes are measured on the VLA
  data obtained as part of the GOODS programme as described in the
  text. Targets are ordered by identifier name (broadly equivalent to
  ordering by date). Transients identified during the 2002-2003
  observing campaign were reported by \citet{2004ApJ...613..200S}.
  Sources from 2004-2005 were classified by and discussed in
  \citet{2008ApJ...681..462D} and \citet{2012ApJ...757...70D}. The ID
  column refers to either the IAU transient name if assigned or the
  identifier assigned to the objects by the GOODS survey team, and
  used in previous literature \citep[e.g.][]{2010MNRAS.tmp..479S}, to
  allow for straightforward comparison. The fifth column gives the rms
  of the local noise in the image for each object, and the sixth
  column the measured signal-to-noise at the supernova location. Six
  objects, marked with an asterisk, are individually
  detected.\label{tab:sne_data}}
\end{table*}

\label{lastpage}


\begin{thebibliography}{99}

\bibitem[\protect\citeauthoryear{Berger et al.}{2003}]{2003ApJ...588...99B} 
Berger E., Cowie L.~L., Kulkarni S.~R., Frail D.~A., Aussel H., Barger 
A.~J., 2003, ApJ, 588, 99 

\bibitem[\protect\citeauthoryear{Castro Cer{\'o}n et 
al.}{2010}]{2010ApJ...721.1919C} Castro Cer{\'o}n J.~M., Micha{\l}owski 
M.~J., Hjorth J., Malesani D., Gorosabel J., Watson D., Fynbo J.~P.~U., 
Morales Calder{\'o}n M., 2010, ApJ, 721, 1919 

\bibitem[\protect\citeauthoryear{Cenko et al.}{2009}]{2009ApJ...693.1484C} 
Cenko S.~B., et al., 2009, ApJ, 693, 1484 

\bibitem[\protect\citeauthoryear{Chandra 
\& Frail}{2012}]{2012ApJ...746..156C} Chandra P., Frail D.~A., 2012, ApJ, 746, 156 

\bibitem[\protect\citeauthoryear{Chapman et 
al.}{2005}]{2005ApJ...622..772C} Chapman S.~C., Blain A.~W., Smail I., 
Ivison R.~J., 2005, ApJ, 622, 772 


\bibitem[\protect\citeauthoryear{Chen et al.}{2009}]{2009ApJ...691..152C} 
Chen H.-W., et al., 2009, ApJ, 691, 152 

\bibitem[\protect\citeauthoryear{Christensen, Hjorth, 
\& Gorosabel}{2004}]{2004A&A...425..913C} Christensen L., Hjorth J., Gorosabel J., 2004, A\&A, 425, 913 


\bibitem[\protect\citeauthoryear{Condon, Cotton, 
\& Broderick}{2002}]{2002AJ....124..675C} Condon J.~J., Cotton W.~D., Broderick J.~J., 2002, AJ, 124, 675 

\bibitem[\protect\citeauthoryear{Coward et al.}{2013}]{2013MNRAS.432.2141C} 
Coward D.~M., Howell E.~J., Branchesi M., Stratta G., Guetta D., Gendre B., 
Macpherson D., 2013, MNRAS, 432, 2141 

\bibitem[\protect\citeauthoryear{Cowie et al.}{1996}]{1996AJ....112..839C} 
Cowie L.~L., Songaila A., Hu E.~M., Cohen J.~G., 1996, AJ, 112, 839 

\bibitem[\protect\citeauthoryear{Cowie et al.}{2004}]{2004ApJ...603L..69C} 
Cowie L.~L., Barger A.~J., Fomalont E.~B., Capak P., 2004, ApJ, 603, L69 

\bibitem[\protect\citeauthoryear{Dahlen, Strolger, 
\& Riess}{2008}]{2008ApJ...681..462D} Dahlen T., Strolger L.-G., Riess A.~G., 2008, ApJ, 681, 462 

\bibitem[\protect\citeauthoryear{Dahlen et al.}{2012}]{2012ApJ...757...70D} 
Dahlen T., Strolger L.-G., Riess A.~G., Mattila S., Kankare E., Mobasher 
B., 2012, ApJ, 757, 70 

\bibitem[\protect\citeauthoryear{Djorgovski et 
al.}{2001}]{2001ApJ...562..654D} Djorgovski S.~G., Frail D.~A., Kulkarni 
S.~R., Bloom J.~S., Odewahn S.~C., Diercks A., 2001, ApJ, 562, 654 


\bibitem[\protect\citeauthoryear{Elliott et 
al.}{2012}]{2012A&A...539A.113E} Elliott J., Greiner J., Khochfar S., Schady P., Johnson J.~L., Rau A., 2012, A\&A, 539, A113 

\bibitem[\protect\citeauthoryear{Evans et al.}{2009}]{2009MNRAS.397.1177E} 
Evans P.~A., et al., 2009, MNRAS, 397, 1177 

\bibitem[\protect\citeauthoryear{Fruchter et 
al.}{2006}]{2006Natur.441..463F} Fruchter A.~S., et al., 2006, Nature, 441, 
463 

\bibitem[\protect\citeauthoryear{Fynbo et al.}{2009}]{2009ApJS..185..526F} 
Fynbo J.~P.~U., et al., 2009, ApJS, 185, 526 

\bibitem[\protect\citeauthoryear{Galama et al.}{1998}]{1998Natur.395..670G} 
Galama T.~J., et al., 1998, Natur, 395, 670 

\bibitem[\protect\citeauthoryear{Gehrels et 
al.}{2004}]{2004ApJ...611.1005G} Gehrels N., et al., 2004, ApJ, 611, 1005 

\bibitem[\protect\citeauthoryear{Giavalisco et 
al.}{2004}]{2004ApJ...600L..93G} Giavalisco M., et al., 2004, ApJ, 600, L93 

\bibitem[\protect\citeauthoryear{Graham 
\& Fruchter}{2013}]{2013ApJ...774..119G} Graham J.~F., Fruchter A.~S., 2013, ApJ, 774, 119 

\bibitem[\protect\citeauthoryear{Greiner et 
al.}{2011}]{2011A&A...526A..30G} Greiner J., et al., 2011, A\&A, 526, A30 

\bibitem[\protect\citeauthoryear{Greiner et 
al.}{2013}]{2013A&A...560A..70G} Greiner J., et al., 2013, A\&A, 560, A70 

\bibitem[\protect\citeauthoryear{Groot et al.}{1998}]{1998ApJ...493L..27G} 
Groot P.~J., et al., 1998, ApJ, 493, L27 

\bibitem[\protect\citeauthoryear{Hatsukade et 
al.}{2012}]{2012ApJ...748..108H} Hatsukade B., Hashimoto T., Ohta K., 
Nakanishi K., Tamura Y., Kohno K., 2012, ApJ, 748, 108 

\bibitem[\protect\citeauthoryear{Hao 
\& Yuan}{2013}]{2013ApJ...772...42H} Hao J.-M., Yuan Y.-F., 2013, ApJ, 772, 42 

\bibitem[\protect\citeauthoryear{Hjorth et al.}{2012}]{2012ApJ...756..187H} 
Hjorth J., et al., 2012, ApJ, 756, 187 

\bibitem[\protect\citeauthoryear{Hopkins et 
al.}{2010}]{2010MNRAS.402.1693H} Hopkins P.~F., Younger J.~D., Hayward 
C.~C., Narayanan D., Hernquist L., 2010, MNRAS, 402, 1693 

\bibitem[\protect\citeauthoryear{Hunt et al.}{2014}]{2014arXiv1402.4006H} 
Hunt L.~K., et al., 2014, arXiv, arXiv:1402.4006 

\bibitem[\protect\citeauthoryear{Jakobsson et 
al.}{2004}]{2004ApJ...617L..21J} Jakobsson P., Hjorth J., Fynbo J.~P.~U., 
Watson D., Pedersen K., Bj{\"o}rnsson G., Gorosabel J., 2004, ApJ, 617, L21 

\bibitem[\protect\citeauthoryear{Kamble et al.}{2014}]{2014arXiv1401.1221K} 
Kamble A., Soderberg A., Berger E., Zauderer A., Chakraborti S., Williams 
P., 2014, arXiv, arXiv:1401.1221 

\bibitem[\protect\citeauthoryear{Kann et al.}{2010}]{2010ApJ...720.1513K} 
Kann D.~A., et al., 2010, ApJ, 720, 1513 


\bibitem[\protect\citeauthoryear{Kelly 
\& Kirshner}{2012}]{2012ApJ...759..107K} Kelly P.~L., Kirshner R.~P., 2012, ApJ, 759, 107 

\bibitem[\protect\citeauthoryear{Kocevski 
\& West}{2011}]{2011ApJ...735L...8K} Kocevski D., West A.~A., 2011, ApJ, 735, L8 


\bibitem[\protect\citeauthoryear{Kocevski, West, 
\& Modjaz}{2009}]{2009ApJ...702..377K} Kocevski D., West A.~A., Modjaz M., 2009, ApJ, 702, 377 

\bibitem[\protect\citeauthoryear{Kohno et al.}{2005}]{2005PASJ...57..147K} 
Kohno K., et al., 2005, PASJ, 57, 147 

\bibitem[\protect\citeauthoryear{Kouveliotou et 
al.}{1993}]{1993ApJ...413L.101K} Kouveliotou C., Meegan C.~A., Fishman 
G.~J., Bhat N.~P., Briggs M.~S., Koshut T.~M., Paciesas W.~S., Pendleton 
G.~N., 1993, ApJ, 413, L101 

\bibitem[\protect\citeauthoryear{Kr{\"u}hler et 
al.}{2011}]{2011A&A...534A.108K} Kr{\"u}hler T., et al., 2011, A\&A, 534, A108 

\bibitem[\protect\citeauthoryear{Kr{\"u}hler et 
al.}{2012}]{2012arXiv1205.4036K} Kr{\"u}hler T., et al., 2012, arXiv, 
arXiv:1205.4036 

\bibitem[\protect\citeauthoryear{Lee, Hwang, 
\& Ko}{2013}]{2013ApJ...774...62L} Lee J.~C., Hwang H.~S., Ko J., 2013, ApJ, 774, 62 

\bibitem[\protect\citeauthoryear{Levesque}{2014}]{2014PASP..126....1L} 
Levesque E.~M., 2014, PASP, 126, 1
 
\bibitem[\protect\citeauthoryear{Levesque et 
al.}{2010a}]{2010AJ....140.1557L} Levesque E.~M., Kewley L.~J., Berger E., 
Zahid H.~J., 2010a, AJ, 140, 1557 

\bibitem[\protect\citeauthoryear{Levesque et 
al.}{2010b}]{2010ApJ...712L..26L} Levesque E.~M., Kewley L.~J., Graham 
J.~F., Fruchter A.~S., 2010b, ApJ, 712, L26 

\bibitem[\protect\citeauthoryear{Mainzer et 
al.}{2011}]{2011ApJ...731...53M} Mainzer A., et al., 2011, ApJ, 731, 53 

\bibitem[\protect\citeauthoryear{Melandri et 
al.}{2012}]{2012MNRAS.421.1265M} Melandri A., et al., 2012, MNRAS, 421, 
1265 
\bibitem[\protect\citeauthoryear{Micha{\l}owski et 
al.}{2012}]{2012ApJ...755...85M} Micha{\l}owski M.~J., et al., 2012, ApJ, 
755, 85 

\bibitem[\protect\citeauthoryear{Micha{\l}owski et 
al.}{2009}]{2009ApJ...693..347M} Micha{\l}owski M.~J., et al., 2009, ApJ, 
693, 347 

\bibitem[\protect\citeauthoryear{Miller et al.}{2008}]{2008ApJS..179..114M} 
Miller N.~A., Fomalont E.~B., Kellermann K.~I., Mainieri V., Norman C., 
Padovani P., Rosati P., Tozzi P., 2008, ApJS, 179, 114 

\bibitem[\protect\citeauthoryear{Moin et al.}{2013}]{2013ApJ...779..105M} 
Moin A., et al., 2013, ApJ, 779, 105 

\bibitem[\protect\citeauthoryear{Morrison et 
al.}{2010}]{2010ApJS..188..178M} Morrison G.~E., Owen F.~N., Dickinson M., 
Ivison R.~J., Ibar E., 2010, ApJS, 188, 178 

\bibitem[\protect\citeauthoryear{Nakagawa et 
al.}{2006}]{2006PASJ...58L..35N} Nakagawa Y.~E., et al., 2006, PASJ, 58, 
L35 

\bibitem[\protect\citeauthoryear{Page et al.}{2005}]{2005MNRAS.363L..76P} 
Page K.~L., et al., 2005, MNRAS, 363, L76 

\bibitem[\protect\citeauthoryear{Pellizza et 
al.}{2006}]{2006A&A...459L...5P} Pellizza L.~J., et al., 2006, A\&A, 459, L5 

\bibitem[\protect\citeauthoryear{Perley et al.}{2013}]{2013ApJ...778..128P} 
Perley D.~A., et al., 2013, ApJ, 778, 128 

\bibitem[\protect\citeauthoryear{Perley et al.}{2008}]{2008ApJ...688..470P} 
Perley D.~A., et al., 2008, ApJ, 688, 470 


\bibitem[\protect\citeauthoryear{Perley et al.}{2007}]{2007GCN..6850....1P} 
Perley D.~A., Chornock R., Bloom J.~S., Fassnacht C., Auger M.~W., 2007, 
GCN, 6850, 1 


\bibitem[\protect\citeauthoryear{Perley 
\& Perley}{2013}]{2013ApJ...778..172P} Perley D.~A., Perley R.~A., 2013, ApJ, 778, 172 


\bibitem[\protect\citeauthoryear{Priddey et 
al.}{2006}]{2006MNRAS.369.1189P} Priddey R.~S. et al, 2006, 
MNRAS, 369, 1189 

\bibitem[\protect\citeauthoryear{Robertson 
\& Ellis}{2012}]{2012ApJ...744...95R} Robertson B.~E., Ellis R.~S., 2012, ApJ, 744, 95 

\bibitem[\protect\citeauthoryear{Rol et al.}{2005}]{2005ApJ...624..868R} 
Rol E., Wijers R.~A.~M.~J., Kouveliotou C., Kaper L., Kaneko Y., 2005, ApJ, 
624, 868 

\bibitem[\protect\citeauthoryear{Salvaterra et 
al.}{2012}]{2012ApJ...749...68S} Salvaterra R., et al., 2012, ApJ, 749, 68 

\bibitem[\protect\citeauthoryear{Sault et al.}{1995}]{1995ASPC...77..433S} 
Sault, R.~J., Teuben, P.~J., 
\& Wright, M.~C.~H.\ 1995, Astronomical Data Analysis Software and Systems IV, 77, 433 

\bibitem[\protect\citeauthoryear{Savaglio, Glazebrook, 
\& LeBorgne}{2009}]{2009ApJ...691..182S} Savaglio\,S.,\,Glazebrook\,K.,\,LeBorgne\,D.,\,2009,\,ApJ,\,691,\,182 


\bibitem[\protect\citeauthoryear{Speagle et 
al.}{2014}]{2014arXiv1405.2041S} Speagle J.~S., Steinhardt C.~L., Capak 
P.~L., Silverman J.~D., 2014, arXiv, arXiv:1405.2041 

\bibitem[\protect\citeauthoryear{Stanway, Davies, 
\& Levan}{2010}]{2010MNRAS.409L..74S} Stanway E.~R., Davies L.~J.~M., Levan A.~J., 2010, MNRAS, 409, L74 


\bibitem[\protect\citeauthoryear{Stanway et 
al.}{2011}]{2011MNRAS.410.1496S} Stanway E.~R., Bremer M.~N., Tanvir N.~R., 
Levan A.~J., Davies L.~J.~M., 2011, MNRAS, 410, 1496 

\bibitem[\protect\citeauthoryear{Strolger et 
al.}{2004}]{2004ApJ...613..200S} Strolger L.-G., et al., 2004, ApJ, 613, 
200 


\bibitem[\protect\citeauthoryear{Svensson et 
al.}{2010}]{2010MNRAS.tmp..479S} Svensson K.~M., Levan A.~J., Tanvir N.~R., 
Fruchter A.~S., Strolger L.-G., 2010, MNRAS, 405, 57 

\bibitem[\protect\citeauthoryear{Svensson et 
al.}{2012}]{2012MNRAS.421...25S} Svensson K.~M., et al., 2012, MNRAS, 421, 
25 

\bibitem[\protect\citeauthoryear{Tanvir et al.}{2004}]{2004MNRAS.352.1073T} 
Tanvir N.~R., et al., 2004, MNRAS, 352, 1073 

\bibitem[\protect\citeauthoryear{Tanvir et al.}{2012}]{2012ApJ...754...46T} 
Tanvir N.~R., et al., 2012, ApJ, 754, 46 

\bibitem[\protect\citeauthoryear{Thoene, Perley, 
\& Bloom}{2007}]{2007GCN..6663....1T} Thoene C.~C., Perley D.~A., Bloom J.~S., 2007, GCN, 6663

\bibitem[\protect\citeauthoryear{Wainwright, Berger, 
\& Penprase}{2007}]{2007ApJ...657..367W} Wainwright C., Berger E., Penprase B.~E., 2007, ApJ, 657, 367 

\bibitem[\protect\citeauthoryear{Wang, Chen, 
\& Huang}{2012}]{2012ApJ...761L..32W} Wang W.-H., Chen H.-W., Huang K.-Y., 2012, ApJ, 761, L32 

\bibitem[\protect\citeauthoryear{Weiler et 
al.}{2002}]{2002ARA&A..40..387W} Weiler K.~W., Panagia N., Montes M.~J., Sramek R.~A., 2002, ARA\&A, 40, 387 

\bibitem[\protect\citeauthoryear{Wirth et al.}{2004}]{2004AJ....127.3121W} 
Wirth G.~D., et al., 2004, AJ, 127, 3121 

\bibitem[\protect\citeauthoryear{Woosley 
\& Heger}{2006}]{2006ApJ...637..914W} Woosley S.~E., Heger A., 2006, ApJ, 637, 914 

\bibitem[\protect\citeauthoryear{Wright}{2006}]{2006PASP..118.1711W} Wright 
E.~L., 2006, PASP, 118, 1711 

\bibitem[\protect\citeauthoryear{Wright et al.}{2010}]{2010AJ....140.1868W} 
Wright E.~L., et al., 2010, AJ, 140, 1868 

\bibitem[\protect\citeauthoryear{Yun \& Carilli}{2002}]{2002ApJ...568...88Y} Yun M.~S., Carilli C.~L., 2002, ApJ, 568, 88 

\bibitem[\protect\citeauthoryear{Zauderer et 
al.}{2013}]{2013ApJ...767..161Z} Zauderer B.~A., et al., 2013, ApJ, 767, 
161 

\end{thebibliography}
\end{document}